\documentstyle[preprint,aps,eqsecnum]{revtex}
\tighten
\begin{document}
\draft

\preprint{\vbox{\hbox{SUSSEX-TH-93/3-4}
\hbox{IMPERIAL/TP/92-93/42}
\hbox{NSF-ITP-93-83}}}

\title{
Evolution of cosmic string configurations
}

\author{Daren Austin and E.J. Copeland}
\address{
School of Mathematical and Physical Sciences,\\
University of Sussex, Brighton BN1 9QH, United Kingdom
}

\author{T.W.B. Kibble}
\address{
Blackett Laboratory, Imperial College,\\
London SW7 2BZ, United Kingdom
}

\date{}
\maketitle

\begin{abstract}
We extend and develop our previous work on the evolution of
a network of cosmic strings.  The new treatment is based on
an analysis of the probability distribution of the end-to-end
distance, or {\it extension}, of a randomly chosen segment of
left-moving string of given length.  The description involves
three distinct length scales: $\xi$, related to the overall
string density, $\bar\xi$, the persistence length along the
string, and $\zeta$, describing the small-scale structure,
which is an important feature of the numerical simulations
that have been done of this problem.  An evolution equation is
derived describing how the distribution develops in time due
to the combined effects of the universal expansion, of
intercommuting and loop formation, and of gravitational
radiation.  With plausible assumptions about the unknown
parameters in the model, we confirm
the conclusions of our previous study,
that if gravitational radiation and small-scale
structure effects are neglected, the two dominant length
scales both scale in proportion to the horizon size.
When the extra  effects are included, we find that while
$\xi$ and $\bar\xi$ grow, $\zeta$ initially does not.
Eventually, however, it does appear to scale, at a much
lower level, due to the effects of gravitational
back-reaction. \end{abstract}

\pacs{98.80.Cq}

\narrowtext

\section{INTRODUCTION}

Cosmic strings are topological defects that may be
formed at a phase transition very early in the history
of the Universe \cite{tk76,ybz}.  Because they are
stable, they may survive to a much later epoch and thus
provide one of the few direct links between the physics
of the very early Universe and recent cosmology.  In
particular, they may play an important role in
generating large-scale structure in the Universe
\cite{av81a,VV91,AS92,rb91}.   Observational tests
of the idea include limits on the gravitational
radiation emitted by collapsing loops and oscillating
strings \cite{HR,SRTR,BB90}, gravitational  lensing of
characteristic form \cite{av81b,av84,jrg} and
predicted anisotropy in the cosmic microwave
background \cite{KS}.

All of these limits depend on our understanding of the
process of evolution of a network of cosmic strings.
One very important question is whether this evolutionary
process leads to a ``scaling'' regime, in which the
characteristic  length scales describing the string
network increase in proportion to the horizon distance
\cite{tk85,db86a,db86b}.   There have been numerical
studies of this problem by at least three different
groups \cite{AT,BB88,AS90}.  There is a wide measure of
agreement between these groups.  All find evidence for
scaling of the large-scale  structure.  There is some
disagreement over the predictions concerning the
small-scale structure, but all groups agree that there
is  a substantial amount of structure on scales much
less than the scale of the long-string network.  It has
certainly become apparent that a simple description in
terms of a single scale is inadequate.

In earlier work \cite{KC,CKA} we  sought to develop an
analytic approach to this problem, to complement the
numerical studies.  In this work, we  identified two
distinct length scales, $\xi$, related to the overall
density of long strings, and $\bar\xi$, the distance over
which the strings are correlated in direction.  It
became apparent, however, that this formalism is
inadequate to represent the small-scale structure
observed in the simulations.  Accordingly, we have
extended our treatment by incorporating a third length
scale, $\zeta$, which describes the structure on the
smallest scales.

This treatment, which we present here, is based on a
probability analysis of the string configuration, in
particular the  probability distribution of the
end-to-end distance, or {\it extension}, ${\bf r}$, of a
randomly chosen segment of  string of length $l$; more
precisely, of {\it left-moving\/} string.  As before, we
use null (characteristic) coordinates on the string
world-sheet.  In flat space, the left- and right-movers
are completely decoupled; the universal expansion
introduces a weak coupling between them.

We should mention that an alternative approach to
obtaining the evolution equations has been proposed by
Embacher, in which a path integral formalism is used to
obtain the probability distribution for the network of
strings in flat space \cite{fe92}.  This promising
approach appears to produce similar results to our own in
the circumstances where comparisons are possible.

In Section II, we review the evolution equations
developed earlier and introduce the fundamental
probability distribution $p[{\bf r}(l)]$ on which our
treatment is based.  We aim to derive an evolution
equation for this quantity, including terms representing
the effects of stretching, gravitational radiation,
intercommuting of long strings and formation of loops.

For all except the very smallest scales, it is
reasonable to  assume that the probability distribution
is Gaussian, characterized by the variance
$\overline{{\bf r}^2} = K(l)$, say.  Our evolution equation
for the probability $p$ effectively reduces to an
equation for this function $K$.  The scales associated
with the large- and small-scale structure on the
strings, $\bar\xi$ and $\zeta$, are defined in terms of $K$,
so their evolution equations follow from that for $K$.

In Section III we analyze the Gaussian Ansatz in more
detail and derive values of various expectation values
that will be  required in the subsequent analysis.  In
particular, we need  a number of joint and conditional
expectation values.

The basic equations describing the effects of
stretching,  intercommuting and loop formation are
reviewed in Section IV. One of the very important things
that emerges from this discussion is that these effects
cannot properly be treated in isolation; they interact
with each other in very complex ways.

Another important lesson is the crucial significance of
the  correlations that develop between left- and
right-movers.  These are discussed in Section V.  In our
previous work \cite{KC,CKA}  we showed that the
stretching process generates such a correlation, and
derived a relation between the mean value $\alpha =
-\overline{{\bf p}{\cdot}{\bf q}}$ (where ${\bf p}$ and ${\bf q}$ are
unit vectors along the left- and right-moving strings)
and the length scale $\bar\xi$.  More recently, however, we
have realised that this is not the only significant
process involved.  In particular, loop formation also
introduces correlations between the ${\bf p}$ and ${\bf q}$
vectors, by preferentially eliminating nearly matching
pairs.  The discussion of this process in Section V is
based on an analysis of the angular probability
distribution of these vectors.

A similar process also introduces angular correlations
on somewhat larger scales.  This is important in
determining the rate of loop formation.  It is treated in
Section VI.  Section VII is devoted to the estimation of
the parameters appearing in the stretching term.  Then in
Section VIII we deal with the effect of
back-reaction from gravitational radiation, which
operates on a very different scale from the other
effects but is ultimately of great importance in the
long-term evolution of the system.

Finally, in Section IX all these terms are brought
together to yield the overall evolution equations for
the three length scales.    The equations
involve four dimensionless functions of the scale
ratios as well as several dimensionless constants.
Before analysing the solutions of the evolution
equations, we study the behaviour of these functions in
various regions of parameter space and try to estimate
the constants.

We then show that so long as the effects of
gravitational radiation are negligible, the system
behaves essentially as predicted by our previous study
\cite{CKA}.  If we start with all three length scales
approximately equal (and somewhat smaller than the
horizon), $\xi$ and later $\bar\xi$ will start to grow and,
under reasonable assumptions, will evolve to a scaling
regime in which they are approximately equal and
proportional to $t$.  The role of the important
parameter $q$ in our previous study is here played by a
ratio between two of the dimensionless functions; it
is, however, no longer a constant.

There is another important parameter, $k$, also
somewhat similar to $q$ (or, more precisely, $q-1$) but
defined in terms of the small-scale structure.  The
value of $k$ is crucial in determining the behaviour of
the third length scale $\zeta$.  If $k$ were large
enough, a complete scaling regime could be reached,
with $\xi$, $\bar\xi$ and $\zeta$ all of comparable
magnitude.  This is not the type of behaviour seen in
the simulations.  A far more likely scenario is that
$k$ is less than the critical value.  Then, while $\xi$
and $\bar\xi \propto t$, the third length scale $\zeta$ does
not grow rapidly, and the ratio $\zeta/\bar\xi$ becomes very
small.  Eventually, when it is of order $\Gamma G\mu$, the
gravitational radiation effect becomes
significant and $\zeta$ starts to grow.  Here
$\Gamma$  is the numerical factor describing the
efficiency of gravitational radiation, estimated to be
of order $10^2$ \cite{av81c,nt84,cjb85,VV85,mh90,QS90}.

We then tackle the important question of whether a
new scaling regime is reached, as various
authors have already suggested \cite{AC91,QP91,da93}, due to
the effects of  gravitational radiation.  It turns out
that the answer to this question depends crucially on
the value of another of the dimensionless
constants we have introduced, $\hat C$, which relates to
the effect of gravitational radiation on small-scale
structure.  Gravitational back-reaction tends to smooth
out the small-scale kinks and thus to make $\zeta$ grow;
$\hat C$ defines the rate at which it does so.  The essential
condition for complete scaling is that $\hat C$ exceeds a
critical value.

We also consider the question of stability and show
that if scaling is achieved it should be stable.

The conclusions are summarized in Section X.  We
discuss in particular a number of remaining open
questions and prospects for future study.

\section{EVOLUTION EQUATIONS}

We shall consider only the case of a $flat$
Robertson-Walker universe, with metric
\begin{equation}
ds^2 = dt^2 - R^2(t)\,d{\bf x}^2
= R^2(\tau)[ d\tau^2 - d{\bf x}^2],
\end{equation}
where $R$ is the Robertson-Walker scale factor,
$\tau = x^0$ is the ``conformal'' time, and
${\bf x}=\{x^k\}$ are
{\it comoving\/} spatial coordinates.

\subsection{The null coordinates}

For completeness, we recall here some of the basic
formalism described in KC \cite{KC}.  It is convenient
to use null (characteristic) coordinates $u,v$ on the
world sheet of the string.  (In flat space, these
are the coordinates $u=t + \sigma,\;
v=t - \sigma$, where $\sigma$ is the length
along the string.)  We denote partial derivatives
with respect to these coordinates by subscripts:
\begin{equation}
x_u = {\partial x\over\partial u},\qquad
x_v = {\partial x\over\partial v}.
\end{equation}
The null condition is
\begin{equation}
x_u^2 = x_v^2 = 0.
\end{equation}
In these coordinates the Nambu-Goto action is
\begin{equation}
I=-\mu \int du\,dv\,R^2(\tau)x_u{\cdot} x_v,
\end{equation}
where $\tau = x^0(u,v)$, and where the dot implies a
scalar product in the Minkowski metric.

In terms of the space-time coordinates $(\tau,{\bf x})$
the equations of motion of the string may be written

 \begin{equation}
\begin{array}{l}
x^0_{uv} = - h_0(x^0_u x^0_v +
{\bf x}_u {\cdot} {\bf x}_v),\\
{\bf x}_{uv} = - h_0(x^0_u {\bf x}_v + {\bf x}_u x^0_v),
\end{array}
\label{eqmot}
 \end{equation}
where
 \begin{equation}
h_\mu = {\partial_\mu R\over R}
= \delta_{\mu 0} {1\over R}{dR\over d\tau}
= \delta_{\mu 0} {dR\over dt}
= \delta_{\mu 0} HR,
\end{equation}
where $H$ is the Hubble parameter.

It is convenient to define the unit vectors
\begin{equation}
{\bf p} = {{\bf x}_u\over x^0_u},\quad
{\bf q} = {{\bf x}_v\over x^0_v},
\label{pqdef}
\end{equation}
which satisfy the equations of motion

\begin{equation}
\begin{array}{l}
{\bf p}_v = - h_0x^0_v({\bf q} - {\bf p}{\bf p}{\cdot}{\bf q}),\\
{\bf q}_u = - h_0x^0_u({\bf p} - {\bf q}{\bf q}{\cdot}{\bf p}).
\end{array}
\label{eqmotq}
\end{equation}

Let us now consider a left-moving segment of string, {\it i.e.},
a segment bounded by two values of the null coordinate
$u$, say $u_1$ and $u_2$.

The total physical extension, or end-to-end distance,
of this segment at a given conformal time, $\tau_0$ say,
is

\begin{eqnarray}
{\bf r}_{\rm tot} &=& R \int_{u_1}^{u_2}
du\left(\partial{\bf x}\over\partial u\right)_\tau\nonumber\\
&=& R \int_{u_1}^{u_2}
du\left[{\bf x}_u + {\bf x}_v\left(
\partial v\over\partial u\right)_\tau\right].
\label{rtotdef}
\end{eqnarray}
In the second term, we can change variable to $v$
and obtain
\begin{equation}
{\bf r}_{\rm tot} = R \int_{u_1}^{u_2} du\,{\bf x}_u
- R \int_{v_2}^{v_1} dv\,{\bf x}_v
\equiv \case{1}/{2}({\bf r}_{\rm l}-{\bf r}_{\rm r}),
\label{rtot}
\end{equation}
say.  (Recall that if $u_1<u_2$ then $v_1>v_2$.)
The two terms in (\ref{rtot}) are the left-moving and
right-moving extensions, respectively.  From now on,
we consider only the left-moving term, and write
\begin{equation}
{\bf r}\equiv{\bf r}_{\rm l}.
\end{equation}

The total physical length of the left-moving segment
(measured along the string) may be defined to be
\begin{equation}
l=2R\int_{u_1}^{u_2} du\,x^0_u\big(u,v(u,\tau_0)\big).
\end{equation}
Note that because the time coordinates at the two
ends have been chosen equal, we necessarily have,
by analogy with (\ref{rtot}),
\begin{equation}
l=2R\int_{u_1}^{u_2} du\,x^0_u
=2R\int_{v_2}^{v_1} dv\,x^0_v.
\end{equation}
In this sense, therefore, the lengths of left- and
right-moving string are exactly equal.

It is useful to note that that the extension ${\bf r}$  can
also be written in terms of the unit vector {\bf p},
defined by (\ref{pqdef}), in the form
\begin{equation}
{\bf r} = \int_0^l dy\,{\bf p}(y),
\end{equation}
where $y$ is a coordinate along the string defined
by
\begin{equation}
dy = 2Rx^0_u\,du.
\label{ydef}
\end{equation}

\subsection{Probability distribution of extension}

Consider a large comoving volume $V$, and let $L$ be the
total length of (left-moving) string within $V$.  It is
convenient to introduce the characteristic inter-string
distance $\xi$ defined by
\begin{equation}
\xi^2 = {V \over L}.
\label{xidef}
\end{equation}

If we
introduce a discretization scale $\delta$, then $L$ must be
thought of as made up of $N$ small segments, each of
length $\delta$, with $N=L/\delta$.  (There will be an
equal length of right-moving string, but
it is convenient to concentrate on one or the other.)

Let us choose a particular length scale $l$ and consider
the probability distribution for the end-to-end distance
(or {\it extension\/}) {\bf r}: $p[{\bf r}(l)] d^3{\bf r}$ is the
probability that a randomly chosen segment of length
$l$ will have extension ${\bf r}$  within the small volume
$d^3{\bf r}$.  Note that in contrast to KC \cite{KC} and
CKA \cite{CKA} we are here using real, rather than
comoving, lengths, {\it i.e.}, in terms of our previous
notation,
\begin{equation}
{\bf r}=R{\bf a},\quad l=Rs.
\label{ra.ls}
\end{equation}

The number of possible starting points within $V$ is
$N$, so the expected number of segments with length
between $l$ and $l+dl$ and extension ${\bf r}$  within
$d^3{\bf r}$ is
\begin{equation}
{L\over\delta}{dl\over\delta}p[{\bf r}(l)]d^3{\bf r}.
\label{segno}
\end{equation}
Because the distribution is highly non-random, there
will be many segments with very similar values of ${\bf r}$
arising from overlapping selections, especially where
at the end-points the orientations happen to be similar.

Strictly speaking, some choices of starting point will
not yield segments entirely within $V$; they will
extend beyond the boundary, but will be matched by a
similar number of segments entering $V$ having
originated outside it.

Our object is, first, to derive an equation for the
rate of change of $p[{\bf r}(l)]$, of the form
\begin{equation}
{\partial p\over\partial t}=
\left({\partial p\over\partial t}\right)_{\rm str.}+
\left({\partial p\over\partial t}\right)_{\rm gr.rad.}+
\left({\partial p\over\partial t}\right)_{\rm l.s.i.}+
\left({\partial p\over\partial t}\right)_{\rm loops},
\label{dpdt}
\end{equation}
and then to determine the nature of its solution.
The various terms on the right represent respectively
the effects of stretching (due to the universal
expansion), of gravitational radiation (back-reaction),
of long-string intercommuting
and of loop production.

The separation between the intercommuting and
loop-production terms is to some extent
arbitrary: there is an upper limit to the size of
a ``large loop''.  However, rather than imposing a
sharp cutoff, we shall aim to count only those loops
that survive the reconnection process.  The reconnection
probability provides a natural cutoff at a scale
determined by the string density.  There is
no need to consider loops larger than this separately
from the long-string network; such loops do not
``know'' that they are closed.  They are rather likely
to suffer reconnection.  Their formation and reconnection
may simply be regarded as instances of long-string
intercommuting.

In addition to the equation (\ref{dpdt}) for $p$, we shall
also need an equation for the rate of change of $L$:
\begin{equation}
{\partial L\over\partial t}=
\left({\partial L\over\partial t}\right)_{\rm str.}+
\left({\partial L\over\partial t}\right)_{\rm gr.rad.}+
\left({\partial L\over\partial t}\right)_{\rm loops}.
\label{dLdt}
\end{equation}
There is no term representing long-string intercommuting,
which has no effect on $L$.

The last term in (\ref{dLdt}), and the last two in
(\ref{dpdt}) are each a
combination of a negative term representing the effect
of removal or destruction of segments and a positive term
representing the corresponding creation of new segments.
It is sometimes easier to consider the change in the
expected number of segments, (\ref{segno}), due to one
of these processes.  It must be remembered that they
affect $L$ as well as $p$, but we note that from the
rate of change of $Lp$ we can easily find those of $L$ and
$p$ separately, by using the normalization condition.
For example,
\begin{equation}
\left({\partial L\over\partial t}\right)_{\rm loops}
= \int d^3{\bf r}
\left({\partial(Lp)\over\partial t}\right)_{\rm loops},
\label{Lp>L}
\end{equation}
and of course
\begin{equation}
\left({\partial p\over\partial  t}\right)_{\rm loops}
= {1\over L} \left({\partial(Lp)\over\partial t}\right)_{\rm loops}
- {p\over L} \left({\partial L\over\partial t}\right)_{\rm loops}.
\label{Lp>p}
\end{equation}

The equations we obtain, not surprisingly, turn out to
be very complicated.  It is unlikely that an exact
solution can be found.  Our aim, however, is to find
approximate solutions valid in special regions of
interest, and in particular to establish whether a
scaling solution exists.  If it does, $p$ should tend
asymptotically to a limiting form
\begin{equation}
p[{\bf r}(l),t] \sim {1\over t^3}
p_{\rm scal.}\left[{{\bf r}\over t}\left(l\over t\right)\right],
\qquad t\to\infty.
\end{equation}

\subsection{The Gaussian Ansatz}

For all except the smallest values of $l$, it is
reasonable to assume that $p$ is a Gaussian,
\begin{equation}
p[{\bf r}(l)] = \left(3\over2\pi K(l)\right)^{3/2}
\exp\left(-{3\over2} {{\bf r}^2\over K(l)}\right).
\label{pGauss}
\end{equation}
Here  $K(l)$ is the mean square extension,
\begin{equation}
K(l) = \overline{{\bf r}^2} =
\int d^3{\bf r}\,{\bf r}^2 p[{\bf r}(l)].
\label{Kdef}
\end{equation}

 From the equation (\ref{dpdt}) for $\partial p/\partial t$
we can derive a corresponding equation for
$\partial K/\partial t$.  We aim to show that its solution
approaches the scaling form
\begin{equation}
K(l,t) \sim t^2
K_{\rm scal.}\left({l\over t}\right),
\qquad t\to\infty.
\end{equation}

For very large values of $l$, $l \gg t$, we expect
the string to behave like a Brownian random walk,
so that $K$ becomes a linear function of $l$,
 \begin{equation}
K(l,t) \sim 2\bar\xi(t) l, \qquad l \gg t,
\label{Kbigl}
 \end{equation}
with persistence length $\bar\xi \propto t$,
but for smaller values of $l$, the variation of $K$
with $l$ will be more rapid, approaching $K \sim l^2$
as $l\to 0$.  (Of course, for such very small values,
the Gaussian approximation breaks down.)

In the next section, we examine the Gaussian Ansatz in
more detail and derive some of its consequences. In
particular, we evaluate various expectation values that
will be needed in the subsequent analysis.  Then
in the following sections, we discuss the various terms
in (\ref{dpdt}) and (\ref{dLdt}).

\section{IMPLICATIONS OF THE GAUSSIAN ANSATZ}

We shall assume that except for very small values of
$l$ the probability distribution of the extension is
a Gaussian, (\ref{pGauss}).  An important function,
introduced in KC \cite{KC}, is the correlation
function
\begin{equation}
f(y) = \overline{{\bf p}(0){\cdot}{\bf p}(y)},
\label{fdef}
\end{equation}
where $y$ is the path-length variable introduced in
(\ref{ydef}).  This function is clearly related to the
variance of {\bf r}.  In fact, we have
\begin{equation}
K(l) \equiv \overline{{\bf r}^2}
= 2\int_0^l dy(l-y)f(y),
\end{equation}
from which it also follows that
\begin{equation}
K'(l) = 2\int_0^l dy\,f(y),
\end{equation}
and
\begin{equation}
K''(l) = 2f(l),
\end{equation}
where the primes denote derivatives with respect
to $l$.

\subsection{An illustrative model}

We shall not make any specific assumption about the
form of $K$ as a function of $l$.  However, it is
useful to have a specific model in mind, to indicate
the kind of behaviour we might expect.

The simplest model, described in KC, is based on the
hypothesis that there is a single scale in the problem:
\begin{equation}
f(l) = e^{-Dl}
\end{equation}
for some constant $D$.  (Note, however, that because
of the change from comoving to real lengths (see
Eq.~(\ref{ra.ls})), $D$ differs from the corresponding
quantity in KC by a factor $R$.)  Then  \begin{equation}
K(l) = {2\over D^2}(e^{-Dl} - 1 + Dl).
\label{K.D}
\end{equation}
The correlation length $\bar\xi$ is defined by
\begin{equation}
\bar\xi = \int_0^\infty dy\,f(y),
\label{xbdef}
\end{equation}
so that for large $l$, $K$ is given by (\ref{Kbigl}).
In the case of the single-scale model,
\begin{equation}
\bar\xi = {1\over D}.
\end{equation}

This model does not describe very well the structure
seen in the simulations.  A better choice might be
a model described by two scales:
\begin{equation}
f(l) = (1-w)e^{-Al} + we^{-Bl},
\label{f2sc}
\end{equation}
where $w$ is a constant in the range $0<w<1$ and we
assume $A\gg B$.  Here the small and large scales
are $1/A$ and $1/B$, respectively.  The expression
for $K$ is similar to (\ref{K.D}):
\begin{equation}
K(l) = {2(1-w)\over A^2}(e^{-Al} - 1 + Al)
+ {2w\over B^2}(e^{-Bl} - 1 + Bl).
\label{K2sc}
\end{equation}
In this case, the correlation length, defined by
(\ref{xbdef}), is
\begin{equation}
\bar\xi = {1-w\over A} + {w\over B} \approx {w\over B}.
\label{xb2sc}
\end{equation}
Note that $\bar\xi$ is dominated by the large scale; the
existence of the small-scale structure does affect it,
via the constant $w$, but the size of the small
scale $1/A$ is more or less immaterial.

It is useful to define another characteristic length
scale, $\zeta$, related to the small-scale structure,
by
\begin{equation}
{1\over\zeta} = - {\partial f(l)\over\partial l}\bigg|_{l=0}.
\label{zedef}
\end{equation}
For the single-scale model, $\zeta=\bar\xi$, but in the
two-scale case,
\begin{equation}
{1\over\zeta} = (1-w)A + wB,
\end{equation}
so
\begin{equation}
\zeta\approx{1\over(1-w)A}.
\end{equation}

For the two-scale model, we may distinguish three
distinct regions, in which the approximate forms of
$K$ (including first-order corrections) are
\begin{equation}
\begin{array}{rll}
(i)&\;l\ll {1\over A}:\quad&
K\approx l^2 - {(1-w)Al^3\over 3};\\
(ii)&\;{1\over A}\ll l \ll {1\over B}:\quad&
K\approx wl^2 - {wBl^3\over 3} + {2(1-w)l\over A};\\
(iii)&\;l\gg {1\over B}:\quad&
K\approx {2wl\over B} - {2w\over B^2} + {2(1-w)l\over A}.
\end{array}
\end{equation}
Leading terms for very small and very large $l$ may
also be written in the forms
\begin{equation}
\begin{array}{rll}
(i)&\;l\ll {1\over A}:\qquad&
K\approx l^2 - {l^3\over 3\zeta};\\
(iii)&\;l\gg {1\over B}:\qquad&
K\approx 2\bar\xi l - {2\bar\xi^2\over w}.
\end{array}
\label{Kapp}
\end{equation}

We emphasize again, however, that this model is
introduced for purely illustrative purposes; we
make no specific assumptions at this stage about
the form of $K$.

\subsection{Higher moments}

The requirement that the probability distribution
is Gaussian means that all its moments are expressible
in terms of the single function $K$.  In particular,
the variance of ${\bf r}^2$, the function
 \begin{equation}
K_{(2)}(l) = \overline{({\bf r}^2)^2}
- (\overline{{\bf r}^2})^2
\label{K2def}
 \end{equation}
is given by
 \begin{equation}
K_{(2)}(l) = \case{2}/{3} K(l)^2.
\label{K2Gauss}
 \end{equation}

The higher cumulants may be found from the cumulant
generating function
\begin{eqnarray}
{\cal K}(z) \equiv \sum_{n=1}^\infty {z^n\over n!}K_{(n)}(l)
&=& \ln\langle e^{z{\bf r}^2}\rangle
\nonumber\\
&=& -\case{3}/{2} \ln\big(1 - \case{2}/{3} zK(l)\big).
\end{eqnarray}

\subsection{Joint probabilities}

In evaluating the various terms in $\partial p/\partial t$,
we shall encounter not only the probability
distribution $p[{\bf r}(l)]$ but also various more
complicated joint probabilities.  We need therefore
to extend our assumptions to cover these.

Consider for example the configuration of two
contiguous segments of lengths $l_1$ and $l_2$
illustrated in Fig.\ \ref{fig1}.

We denote the joint probability of extensions ${\bf r}_1$
and ${\bf r}_2$ within small volumes $d^3{\bf r}_1$ and
$d^3{\bf r}_2$ by
\begin{equation}
p[{\bf r}_1(l_1),{\bf r}_2(l_2)]\,d^3{\bf r}_1\,d^3{\bf r}_2.
\label{jtprob}
\end{equation}

For a Brownian process, this probability factorizes
into the product $p[{\bf r}_1(l_1)].p[{\bf r}_2(l_2)]$, but
this is possible only if $K(l)$ is a linear function
of $l$, since it implies that $\overline{{\bf r}^2}
= \overline{{\bf r}_1^2} + \overline{{\bf r}_2^2}$.  In
the general case, where this is not true,
$\overline{{\bf r}_1{\cdot}{\bf r}_2}$ is non-zero.  In fact,
\begin{equation}
K(l_1,l_2) \equiv \overline{{\bf r}_1{\cdot}{\bf r}_2}
= \case{1}/{2}[K(l) - K(l_1) - K(l_2)],
\label{K12def}
\end{equation}
where of course
\begin{equation}
l = l_1 + l_2.
\end{equation}

It is interesting to examine the limiting forms of the
expression (\ref{K12def}) in our two-scale model.  The
leading terms in the three regions are:
\begin{equation}
\begin{array}{rll}
(i)&\;l_1,l_2\ll {1\over A}:\quad&
K(l_1,l_2)\approx l_1l_2;\\
(ii)&\;{1\over A}\ll l_1,l_2\ll {1\over B}:\quad&
K(l_1,l_2)\approx wl_1l_2;\\
(iii)&\;l_1,l_2\gg {1\over B}:\quad&
K(l_1,l_2)\approx {w\over B^2}.
\end{array}
\end{equation}
The constant value as $l_1$ and $l_2$ approach infinity
is noteworthy.

It is useful to note that a similar formula to
(\ref{K12def}) holds for the mean value of the scalar
product of overlapping segments; for example, in the
configuration of Fig.\ \ref{fig1},
\begin{equation}
\overline{{\bf r}_1{\cdot}{\bf r}} = \case{1}/{2}
[K(l) + K(l_1) - K(l_2)].
\label{K10ov}
\end{equation}
This expression of course is non-zero even for linear
$K$.

In line with the Gaussian Ansatz for $p[{\bf r}(l)]$,
we shall assume that the joint probability
(\ref{jtprob}) is also Gaussian, except of course
for very small values of $l_1$ or $l_2$.  Its form is
then completely determined by the covariance matrix
\begin{equation}
{\bf K} \equiv \left[
\begin{array}{cc}
\overline{{\bf r}_1^2}&\overline{{\bf r}_1{\cdot}{\bf r}_2}\\
\overline{{\bf r}_1{\cdot}{\bf r}_2}&\overline{{\bf r}_2^2}
\end{array}
\right]=\biggl[
\begin{array}{cc}
K(l_1)&K(l_1,l_2)\\
K(l_1,l_2)&K(l_2)
\end{array}
\biggr].
\end{equation}
The distribution may be written
\begin{equation}
p[{\bf r}_1,{\bf r}_2] = \left(3\over2\pi\right)^3
{1\over(\det{\bf K})^{3/2}}\exp[-\case{3}/{2}
\tilde{\bf R}{\cdot}{\bf K}^{-1}{\bf R}],
\label{p12}
\end{equation}
where
\begin{equation}
{\bf R} = \biggl[
\begin{array}{c}
{\bf r}_1\\{\bf r}_2
\end{array}
\biggr].
\end{equation}
The determinant in (\ref{p12}) may be written, using
(\ref{K12def}), as a symmetric function of the three
variables
\begin{equation}
K=K(l),\quad K_1=K(l_1),\quad K_2=K(l_2),
\end{equation}
namely
\begin{equation}
\det{\bf K} = \case{1}/{4}[2KK_1+2KK_2+2K_1K_2
-K^2-K_1^2-K_2^2].
\end{equation}
We note that the determinant is automatically positive
provided that $K$ increases as a function of $l$
faster than linearly but less than quadratically, so
that
\begin{equation}
K_1+K_2 < K < (\sqrt{K_1} + \sqrt{K_2})^2.
\end{equation}
This is assured if $f(l)$ is a positive, monotonically
decreasing function.

It is useful to note that the joint probability
(\ref{jtprob}) could equally well be written in
terms of the variables ${\bf r}_1$  and ${\bf r}$, say, rather
than ${\bf r}_1$ and ${\bf r}_2$, provided of course that
${\bf K}$  were replaced by the appropriate covariance matrix,
\begin{equation}
{\bf K}\,\to\,\Biggl[
\begin{array}{cc}
\overline{{\bf r}_1^2}&\overline{{\bf r}_1{\cdot}{\bf r}}\\
\overline{{\bf r}_1{\cdot}{\bf r}}&\overline{{\bf r}^2}
\end{array}
\Biggr].
\label{K01}
\end{equation}

\subsection{Conditional expectation values}

In later sections, we shall need various conditional
expectation values, for example the expectation value
of the extension ${\bf r}_1$ of the $l_1$ segment in
Fig.\ \ref{fig1} for a given value of the overall extension
${\bf r}$  of the composite segment of length $l$.  Let
us denote this conditional averaging, over the
ensemble of segments with given values of ${\bf r}$  and
$l$, by angle brackets.

 From the joint probability (\ref{p12}), it is
straightforward to evaluate this conditional expectation
value,
 \begin{equation}
\langle{\bf r}_1\rangle = \int d^3{\bf r}_1\,
{\bf r}_1 p[{\bf r}_1|{\bf r}], \qquad
p[{\bf r}_1|{\bf r}] = {p[{\bf r}_1,{\bf r}]\over p[{\bf r}]}.
 \end{equation}
We find
 \begin{eqnarray}
\langle{\bf r}_1\rangle &=&
{\overline{{\bf r}_1{\cdot}{\bf r}}\over
\overline{{\bf r}^2}}{\bf r}
\label{r1exp1}\\
&=& {K(l) + K(l_1) - K(l_2)\over 2K(l)}{\bf r}.
\label{r1exp}
 \end{eqnarray}
This is not at all surprising: a selected portion of
a segment of extension ${\bf r}$  is obviously more likely
to have its own extension ${\bf r}_1$ in the same
direction as {\bf r}.

For large values of $l$ and $l_1$, for which $K$ is
linear, this yields, as one might expect
 \begin{equation}
\langle{\bf r}_1\rangle \approx
{l_1\over l}{\bf r} \qquad({\rm large}\;l_1,l).
 \end{equation}

 From (\ref{r1exp1}) it follows of course that
 \begin{equation}
\langle{\bf r}{\cdot}{\bf r}_1\rangle =
{\overline{{\bf r}{\cdot}{\bf r}_1}\over
\overline{{\bf r}^2}}{\bf r}^2.
 \end{equation}

Less obvious, but also useful, is the conditional
expectation value of ${\bf r}_1^2$.  A straightforward
calculation yields
\begin{equation}
\langle{\bf r}_1^2\rangle =
{\overline{{\bf r}^2}\,\overline{{\bf r}_1^2}
-(\overline{{\bf r}{\cdot}{\bf r}_1})^2\over
\overline{{\bf r}^2}}+
{(\overline{{\bf r}{\cdot}{\bf r}_1})^2\over
(\overline{{\bf r}^2})^2} {\bf r}^2.
\label{r1sqexp}
\end{equation}
 From (\ref{r1exp}) and (\ref{r1sqexp}) we obtain the
conditional variance of ${\bf r}_1$:
\begin{equation}
\langle{\bf r}_1^2\rangle - (\langle{\bf r}_1\rangle)^2 =
{\overline{{\bf r}^2}\,\overline{{\bf r}_1^2}
-(\overline{{\bf r}{\cdot}{\bf r}_1})^2\over
\overline{{\bf r}^2}}.
\label{r1var}
\end{equation}
It is remarkable that this conditional variance is
in fact independent of the actual value of {\bf r}.  Note
however that for {\it any\/} given value of {\bf r},
it is less than the {\it unconditional\/} variance
$\overline{{\bf r}_1^2}$.

A useful corollary of (\ref{r1exp}) yields the
conditional expectation value of the unit vector
{\bf p}.  We can write
\begin{equation}
{\bf p}(l_1) = {\partial {\bf r}_1\over\partial l_1},
\label{p1def}
\end{equation}
whence
\begin{equation}
\overline{{\bf r}{\cdot}{\bf p}(l_1)} =
\case{1}/{2}[K'(l_1) + K'(l-l_1)],
\label{rpexp}
\end{equation}
and
\begin{equation}
\langle {\bf p}(l_1)\rangle =
{\overline{{\bf r}{\cdot}{\bf p}(l_1)}\over
\overline{{\bf r}^2}}{\bf r} =
{K'(l_1) + K'(l-l_1)\over2K(l)}{\bf r}.
\label{pexp}
\end{equation}
For large values of $l$ and $l_1$, this reduces to
\begin{equation}
\langle {\bf p}(l_1)\rangle \approx {{\bf r}\over l}\quad
({\rm large}\; l_1,l),
\end{equation}
again as one might expect.

It is useful to note that these results, in particular
(\ref{pexp}), continue to hold for {\it negative\/}
values of $l_1$, provided that for negative $l$,
$K(l)$ is interpreted as meaning $K(|l|)$.  In other
words, because the correlation of direction extends
over a finite distance, a small segment close to but
outside our chosen segment will still be correlated
with it, though clearly less strongly than if it
were inside.  Of course, as $l_1 \to -\infty$,
$K'(l_1) \to -2\bar\xi$, while $K'(l-l_1) \to 2\bar\xi$, so
eventually $\langle {\bf p}(l_1)\rangle$ does approach
zero.

\subsection{Triple joint probabilities}

We shall also need to consider more complex joint
probabilities, such as the triple joint probability
of the configuration shown in Fig.\ \ref{fig2}.

This may be written as
 \begin{equation}
p[{\bf r}_0,{\bf r}_1,{\bf r}] = \left(3\over2\pi\right)^{9/2}
{1\over(\det{\bf K})^{3/2}}\exp[-\case{3}/{2}
\tilde{\bf R}{\cdot}{\bf K}^{-1}{\bf R}],
\label{p120}
 \end{equation}
where now
 \begin{equation}
{\bf R} = \left[
\begin{array}{c}
{\bf r}_0\\{\bf r}_1\\{\bf r}
\end{array}
\right],\quad
{\bf K} = \left[
\begin{array}{ccc}
\overline{{\bf r}_0^2}&\;\overline{{\bf r}_0{\cdot}{\bf r}_1}\;
&\overline{{\bf r}_0{\cdot}{\bf r}}\\
\overline{{\bf r}_0{\cdot}{\bf r}_1}&\overline{{\bf r}_1^2}
&\overline{{\bf r}_1{\cdot}{\bf r}}\\
\overline{{\bf r}_0{\cdot}{\bf r}}&
\;\overline{{\bf r}_1{\cdot}{\bf r}}\;&\overline{{\bf r}^2}
\end{array}
\right].
 \end{equation}

Integrating over all values of ${\bf r}_0$ yields a joint
probability distribution for ${\bf r}_1$ and ${\bf r}$  identical
in form to the previous one, save for the fact that
the expression for $\overline{{\bf r}_1{\cdot}{\bf r}}$
is different, namely
 \begin{equation}
\overline{{\bf r}_1{\cdot}{\bf r}} = \case{1}/{2}
[K(l_0+l_1) + K(l-l_0) - K(l_0) - K(l-l_0-l_1)].
\label{r10exp}
 \end{equation}
With this change, the previous equations remain valid.

Another useful result can be obtained from (\ref{p120}).
We can find an expression for the conditional average
of a scalar product of unit vectors,
$\langle {\bf p}_0{\cdot}{\bf p}_2\rangle$, where
${\bf p}_0 = {\bf p}(l_0)$ and ${\bf p}_2 = {\bf p}(l_0+l_1)$,
by using analogues of (\ref{p1def}), namely
 \begin{equation}
{\bf p}_0 = {\partial {\bf r}_0\over\partial l_0},\quad
{\bf p}_2 = {\partial {\bf r}_2\over\partial l_2},
 \end{equation}
with $l_2 = l - l_0 - l_1$.  We find
 \begin{eqnarray}
&&\langle{\bf r}_0{\cdot}{\bf r}_2\rangle =
\overline{{\bf r}_0{\cdot}{\bf r}_2} +
{\overline{{\bf r}_0{\cdot}{\bf r}}\,
\overline{{\bf r}_2{\cdot}{\bf r}}\over
(\overline{{\bf r}^2})^2}({\bf r}^2 - \overline{{\bf r}^2})
\nonumber\\
&&\quad= \case{1}/{2}[K(l) + K(l-l_0-l_2) -
K(l-l_0) - K(l-l_2)]\nonumber\\
&&\quad+ \case{1}/{4}[K(l) + K(l_0) - K(l-l_0)]
[K(l) + K(l_2) - K(l-l_2)]
{{\bf r}^2 - K(l)\over K(l)^2}.
 \end{eqnarray}
Hence, differentiating with respect to both $l_0$ and
$l_2$, we get
 \begin{eqnarray}
&&\langle {\bf p}_0{\cdot}{\bf p}_2\rangle =
\case{1}/{2} K''(l-l_0-l_2)\nonumber\\
&&\quad+ \case{1}/{4} [K'(l_0) + K'(l-l_0)]
[K'(l_2) + K'(l-l_2)]
{{\bf r}^2 - K(l)\over K(l)^2}.
\label{ppexp}
 \end{eqnarray}
Comparing with (\ref{pexp}), we see that the conditional
covariance function of ${\bf p}_0$ and ${\bf p}_2$ is again
independent of ${\bf r}$ (but smaller than the unconditional
value):
 \begin{eqnarray}
\langle {\bf p}_0{\cdot}{\bf p}_2\rangle
&-& \langle {\bf p}_0\rangle {\cdot}
\langle{\bf p}_2\rangle =
\case{1}/{2} K''(l_1)\nonumber\\
&-& {1\over4K(l)}[K'(l_0) + K'(l-l_0)]
[K'(l_2) + K'(l-l_2)].
 \end{eqnarray}

Note that as before these equations continue to hold
even when either $l_0$ or $l_2$ is negative.

\subsection{Small-$l$ behaviour}
\label{sec-smalll}

The Gaussian approximation obviously breaks down for
very small values of $l$.  To obtain information about
the time evolution of the smallest-scale structures,
we need to have approximate formulae that can be used
in that region too.

For very small $l$, the expectation value of
${\bf r}^2$ takes the form given in (\ref{Kapp}({\it i})),
\begin{equation}
K\approx l^2 - {l^3\over 3\zeta}.
\label{Ksmall}
\end{equation}

We shall also need to consider the probability
distribution for a small segment of length $l_1$
within a larger segment of length $l$ and extension
{\bf r}.  From (\ref{pexp}), it follows that the
{\it direction\/} of ${\bf r}_1$ is correlated with that
of {\bf r}.  However, the length is not; the expression
(\ref{r1sqexp}) does {\it not\/} hold in the limit of small
$l_1$.  In fact, in that limit the length is essentially
fixed; the probability distribution for ${\bf r}_1$ is
concentrated in a thin shell near $|{\bf r}_1| = l_1$.
Therefore,
\begin{equation}
\langle{\bf r}_1^2\rangle \approx
\overline{{\bf r}_1^2} \approx l_1^2,
\qquad l_1\to 0.
\label{r1sqsmall}
\end{equation}

Another important difference from the Gaussian case
concerns the variance of ${\bf r}^2$, which is no longer given
by (\ref{K2Gauss}).  In fact, for small $l$,
the leading term in $K_{(2)}$ is clearly of order $l^5$.
To obtain a more specific result, consider for
example a model in which small-angle kinks are
randomly distributed on the string, with $D$ kinks
per unit length, and suppose the kink angles are
distributed according to a (two-dimensional)
distribution with (small) variance $\overline{\theta^2}$.
Then for values of $l$ such that $Dl \ll 1$, we find
\begin{equation}
K(l) = l^2 - \case{1}/{3}
Dl^3(1-\overline{\cos\theta})
\approx l^2 - \case{1}/{6}Dl^3\overline{\theta^2}.
\end{equation}
The characteristic scale $\zeta$ of the
small-scale structure, defined by (\ref{zedef}), is
\begin{equation}
\zeta \approx {2\over\overline{\theta^2}D}.
\label{zesmall}
\end{equation}
Similarly, we obtain
\begin{equation}
K_{(2)}(l) \approx \case{1}/{30}Dl^5\overline{\theta^4},
\qquad l\to 0.
\label{K2small}
\end{equation}
The ratio $K_{(2)}/K^2$, which is ${2\over3}$ in the
Gaussian case, becomes
\begin{equation}
{K_{(2)}(l)\over K(l)^2} \approx
{\overline{\theta^4}\over30\overline{\theta^2}}
{l\over\zeta}\to 0 \; {\rm as}\;l\to0.
\label{K2l0}
\end{equation}
In particular, if the distribution of kink angles is
Gaussian, then
\begin{equation}
{K_{(2)}(l)\over K(l)^2} \approx
{\overline{\theta^2}\over15} {l\over\zeta}.
\label{K2l0G}
\end{equation}

\section{DERIVATION OF BASIC RATE EQUATIONS}

The evolution of the system of strings is a complicated
process.  The mechanisms represented by the various
terms in (\ref{dpdt}) do  not act independently.  At
least to a first approximation, gravitational radiation
is separable from the others, because the dominant
scale involved is much smaller.  We shall therefore
postpone its discussion.  As we shall see, however,
the remaining three act in a complex synergy.

We begin this section by deriving the basic equations
for these processes,
reviewing for completeness the discussion of KC and
CKA \cite{KC,CKA}.

\subsection{Rates of change of length and extension}

 From the equations of motion (\ref{eqmot}), we can
derive expressions for the expected rates of change
of the length $l$ and extension ${\bf r}$  of a chosen
segment, expressing ${\dot l}$ and $\dot{\bf r}$ as
functions of $l$ and {\bf r}.  Here, unlike KC, we use
dots to denote derivatives with respect to real
time; in particular, ${\dot l}$ stands for
the average value, $\langle dl/dt \rangle$, of $dl/dt$
over the ensemble of segments of given length and
extension.

Similarly, we can derive an expression for
$(\dot L)_{\rm str.}$, the contribution of stretching
to the rate of increase in the overall length of
string within a given large comoving volume.

It is very important to note that there are consistency
requirements.  As before, let us denote by an overbar
the average value of any function of ${\bf r}$  over the
probability distribution $p[{\bf r}(l)]$ for given
length $l$, {\it e.g.}
\begin{equation}
\overline{{\dot l}} = \int d^3{\bf r}\,{\dot l} p[{\bf r}(l)].
\end{equation}
The lengths $l$ of different segments will stretch
by different amounts, depending on the value of ${\bf r}$
and other factors.  However, the entire string can
be chopped up conceptually
into segments of any prescribed length
$l$, and its overall growth is obviously independent of
$l$.  Thus for every value of $l$ we must have
\begin{equation}
{\overline{{\dot l}}\over l} =
{(\dot L)_{\rm str.}\over L}.
\label{consist}
\end{equation}

Deriving the expression for
$(\partial p/\partial t)_{\rm str.}$ from ${\dot l}$
and $\dot{\bf r}$ is actually quite subtle, because
of the dependence of ${\dot l}$ on {\bf r}; a sample of
segments initially of equal lengths does not remain so.

Consider a small time interval $dt$ during which the
expected changes of $l$ and ${\bf r}$  are
\begin{equation}
l \to l' = l + {\dot l} dt,\qquad
{\bf r} \to {\bf r}' = {\bf r} + \dot{\bf r} dt.
\label{ll'rr'}
\end{equation}
The total length will also change of course, according
to
\begin{equation}
L \to L' = L + (\dot L)_{\rm str.} dt.
\end{equation}

Now suppose that within $V$ we select segments at
random by choosing independently a random starting
point and a random length $l$, chosen from a uniform
distribution from 0 up to a large upper limit, say
$L$.  The changes of length and extension of
the chosen segments over a short time interval $dt$
will vary randomly, with expectation values given
by (\ref{ll'rr'}).  However, because of the consistency
requirement (\ref{consist}), the final distribution
will still be {\it uniformly distributed\/} in $l'$ (at
least so long as $l\ll L$, so that the upper cutoff of
lengths is irrelevant).  Hence we have the important
equality
\begin{equation}
p[{\bf r}'(l'),t']{dl'\over L'}d^3{\bf r}' =
p[{\bf r}(l),t]{dl\over L}d^3{\bf r}.
\end{equation}
Now
\begin{equation}
dl' = \left(1 + dt{\partial{\dot l}\over\partial l}\right)dl
 \end{equation}
and
 \begin{equation}
d^3{\bf r}' = \left(1 + dt{\partial\over\partial{\bf r}}
{\cdot}\dot{\bf r}\right) d^3{\bf r}.
 \end{equation}
Hence it follows that
 \begin{equation}
\left(\partial p\over\partial t\right)_{\rm str.} =
- {\partial\over\partial l}({\dot l} p)
- {\partial\over\partial{\bf r}}{\cdot}(\dot{\bf r} p)
+ {(\dot L)_{\rm str.}\over L}p.
\label{pdstr}
 \end{equation}
This is consistent with the normalization condition
because, from (\ref{consist}),
 \begin{equation}
{\partial\over\partial l}\overline{{\dot l}} =
{(\dot L)_{\rm str.}\over L}.
 \end{equation}

As before, let us consider a left-moving segment of
string, defined by the inequalities $u_1 < u < u_2$.
The length of our chosen segment at time
$t$ is
\begin{equation}
l(t) = 2R \int_{u_1}^{u_2} du\,x^0_u(u, v(u,t)),
\end{equation}
while its extension is
\begin{equation}
{\bf r}(t) = 2R \int_{u_1}^{u_2} du\,{\bf x}_u(u, v(u,t))
\end{equation}

To find the rates of change, we use
 \begin{equation}
\left(\partial v\over\partial t\right)_u =
{1\over R}\left(\partial v\over\partial\tau\right)_u =
{1\over R}\left(\partial \tau\over\partial v\right)_u^{-1}
= {1\over Rx^0_v}.
 \end{equation}
Hence from the equations of motion (\ref{eqmot})
 \begin{eqnarray}
{dl\over dt} &=& {\dot R\over R}l +
2 \int_{u_1}^{u_2} du\,x^0_{uv}{1\over x^0_v}
\nonumber\\
&=& {\dot R\over R}l -
2 \dot R\int_{u_1}^{u_2} du\,x^0_u(1+{\bf p}{\cdot}{\bf q})
\nonumber\\
&=& - 2 \dot R\int_{u_1}^{u_2} du\,x^0_u{\bf p}{\cdot}{\bf q}.
 \end{eqnarray}

As before, we denote averaging over the ensemble of
segments with given values of $l$ and ${\bf r}$  by angle
brackets.  We then have
 \begin{equation}
{\dot l} =  - 2 \dot R\int_{u_1}^{u_2} du\,
\langle x^0_u{\bf p}{\cdot}{\bf q}\rangle
\equiv \alpha({\bf r},l){\dot R\over R}l,
\label{aldef}
 \end{equation}
say.  Note that in order to satisfy the consistency
requirement (\ref{consist}), the average value of
$\alpha$ over the {\bf r}-distribution must be a constant
$\bar\alpha$, independent of $l$.  We must have
 \begin{equation}
(\dot L)_{\rm str.} = \bar\alpha{\dot R\over R}L.
\label{Ldstr}
 \end{equation}
Thus $\bar\alpha$ may be identified with the constant
$\alpha$ of KC \cite{KC} and CKA \cite{CKA}.    More
generally, we shall find later that $\alpha$ has the form
 \begin{equation}
\alpha({\bf r},l) = \bar\alpha + \hat\alpha (l)
{{\bf r}^2 - K(l)\over K(l)}.
\label{alsplit}
 \end{equation}

Similarly, we obtain
 \begin{equation}
{d{\bf r}\over dt} = {\dot R\over R}{\bf r} -
2 \dot R\int_{u_1}^{u_2} du\,x^0_u({\bf p}+{\bf q})
 \end{equation}
and hence
 \begin{equation}
\dot{\bf r} =  - 2 \dot R\int_{u_1}^{u_2} du\,
\langle x^0_u{\bf q}\rangle
\equiv \beta({\bf r},l){\dot R\over R}{\bf r}.
\label{betadef}
 \end{equation}

The parameters $\alpha$ and $\beta$ are related but
distinct.  In particular, there is no special
condition on the mean value of $\beta$.   It will
emerge later that $\beta$ may be taken to be a function
of $l$ alone, independent of ${\bf r}$.

It is convenient to rewrite (\ref{aldef}) and
(\ref{betadef}) in terms of the path-length variable
$y$ introduced in (\ref{ydef}).  We then have
 \begin{equation}
\alpha l = - \int_0^l dy\,\langle
{\bf p}{\cdot}{\bf q} \rangle
\label{aldef1}
 \end{equation}
and
 \begin{equation}
\beta {\bf r} =  - \int_0^l dy\,\langle {\bf q} \rangle.
\label{betadef1}
\end{equation}

We shall return to the computation of these
averages in the next Section.

 Substituting $\dot l$ and $\dot{\bf r}$ into
(\ref{pdstr}), and using (\ref{alsplit}), we thus find
 \begin{equation}
\left(\partial p\over\partial t\right)_{\rm str.} =
- H\bar\alpha l{\partial p\over\partial l}
- H{\partial\over\partial l}\left(l\hat\alpha (l)
{{\bf r}^2-K\over K}p\right)
- H\beta(l){\partial\over\partial{\bf r}}{\cdot}({\bf r} p),
\label{pdstr2}
 \end{equation}
where as before $H$ is the Hubble parameter,
$H = \dot R/R$.

Taking a moment of (\ref{pdstr2}), we find for the rate
of change of $K$,
 \begin{eqnarray}
\left(\partial K\over\partial t\right)_{\rm str.} &=&
- H{\partial\over\partial l} \left(
\bar\alpha lK + \hat\alpha  l{K_{(2)}\over K} \right)
+ 2H\beta K + H\bar\alpha K
\nonumber\\
&=& H \biggl[2\beta K - \bar\alpha l{\partial K\over\partial l}
- {\partial\over\partial l}\left(\hat\alpha  l
{K_{(2)}\over K}\right) \biggr].
\label{Kdstr}
 \end{eqnarray}

\subsection{Intercommuting probability}
\label{sec-lsi}
Next we review briefly the derivation of the
intercommuting probability given
in KC \cite{KC}.  This also provides an opportunity to
refine the argument.

Consider a large volume $V\/$, which, according to
(\ref{xidef}), contains a length $L=V/\xi^2$ of long
string.  In other words, each volume $\xi^3$ contains
on average a length $\xi$ of string.  The model
introduced in KC was to regard this string as formed
of $N$ independently moving straight segments, each
of length $\xi$, where
\begin{equation}
N = {L\over\xi} = {V\over\xi^3}.
\label{Ndef}
\end{equation}

Choose any pair of these segments.  For simplicity,
assume that at the relevant time, the spatial coordinate
$\sigma$ along the string is chosen to coincide with
the variable $y$ which measures length along the
string.  The two segments will intersect
at some time during a short interval $\delta t$ if
there is a solution to
\begin{equation}
{\bf x}_{01} + y_1{\bf x}_1' + t\dot{\bf x}_1 =
{\bf x}_{02} + y_2{\bf x}_2' + t\dot{\bf x}_2,
\end{equation}
with
\begin{equation}
0<y_1<\xi,\qquad 0<y_2<\xi,\qquad 0<t<\delta t.
\end{equation}

Equivalently, if the starting point of one of
the segments, ${\bf x}_{01}$, is fixed, intercommuting will
occur if the other starting point, ${\bf x}_{02}$, lies
within a small volume
\begin{equation}
\delta V = \xi^2 \delta t |{\bf x}_1'\wedge{\bf x}_2'{\cdot}
(\dot{\bf x}_1 - \dot{\bf x}_2)|.
\end{equation}
(The additional factor of $1\over4$ appearing in
KC, equation (4.26), was an error.)  The probability of
intercommuting between this pair of segments is
$\delta V/V$.  To obtain the probability that a chosen
segment undergoes intercommuting, we multiply by the
number of segments, $N$, given by (\ref{Ndef}).  Thus the
probability that a  string segment of length $l$ will
undergo intercommuting during a time interval $dt$ is
\begin{equation}
\chi{l\,dt\over\xi^2},
\end{equation}
where $\chi$ is the average value of the scalar triple
product $|{\bf x}_1'\wedge{\bf x}_2'{\cdot}
(\dot{\bf x}_1 - \dot{\bf x}_2)|$.

The {r.m.s.} values
of $|{\bf x}'|$ and $|\dot{\bf x}|$ are
$\sqrt{(1+\alpha)/2}$ and $\sqrt{(1-\alpha)/2}$,
respectively.  One way of estimating $\chi$ is to use
these as typical values.  Then averaging over all
angles, maintaining the orthogonality of ${\bf x}'$ and
$\dot{\bf x}$, introduces a factor of $2/\pi$.  Thus
typically
 \begin{equation}
\chi = {1+\alpha\over\pi}
\sqrt{1-\alpha\over2} \approx 0.24.
\label{chi1}
\end{equation}
Although $\chi$ therefore has a weak $\alpha$
dependence, this is probably not sufficiently important
to make it necessary to use an {\bf r}-dependent value.

Another way of estimating $\chi$ is to rewrite the
scalar triple product in terms of ${\bf p}$ and ${\bf q}$
vectors, as
 \begin{equation}
\case{1}/{4} |{\bf p}_1\wedge{\bf p}_2{\cdot}
({\bf q}_1 - {\bf q}_2) + ({\bf p}_1 - {\bf p}_2)
{\cdot}{\bf q}_1\wedge{\bf q}_2|.
\label{sctrpr}
 \end{equation}
Averaging over all directions of the  ${\bf p}$ and ${\bf q}$
vectors, assuming that they are independently and
isotropically distributed, yields
\begin{equation}
\chi = {2\pi\over 35} \approx 0.18.
\label{chi2}
\end{equation}
(It is easy to see why this result should be smaller
than our first estimate (\ref{chi1}).  Using typical
average values of the magnitudes of ${\bf x}'$ and $\dot{\bf x}$
ignores the anticorrelation between them; including it
would tend to reduce the estimate.)  We can
improve the estimate (\ref{chi2}) by allowing for the
correlation between the  ${\bf p}$ and ${\bf q}$ vectors,
corresponding to the non-zero value of $\alpha$.  This gives
a corrected value
 \begin{equation} \chi = {2\pi\over 35}+
{4\pi\alpha\over 105} \approx 0.20.
\label{chi3}
 \end{equation}
This is probably the
best estimate we can achieve.

It might be argued that the model used here, assuming
straight segments of length $\xi$, is inaccurate.  The
length scale on which strings are roughly straight is
not $\xi$ but $\bar\xi$.  A better approximation might be
to assume that the string is composed of
straight segments of length $\bar\xi$. However, this
actually makes no difference to the final answer.  The
small volume $\delta V$ is then proportional to
$\bar\xi^2 \delta t$, while $N$ becomes $N = V/\xi^2\bar\xi$,
so the probability of intercommuting is still proportional
to the length of the segment, in this case $\bar\xi$.

The assumption that the segments are uncorrelated may
seem rather dubious, particularly in the case where
$\bar\xi$ is small compared to $\xi$.  Within each volume
$\xi^3$, there would be a length $\xi$ of string,
consisting not of a single straight section but
of several segments of length $\bar\xi$.  To treat these
as moving independently is clearly to underestimate
the number of intercommutings.  However, the
underestimate refers to intercommutings between
{\it nearby\/} pairs, which will of course lead to loop
formation and are therefore included elsewhere.

One might also be concerned about the effects of
small-scale structure on the strings.  Segments that
are roughly straight but kinky on a small scale would
seem to offer a smaller cross-section.  But in fact,
this too is irrelevant to the probability of long-string
intercommuting, though very relevant to loop formation.
Consider two kinky segments of string, each of length
$\bar\xi$.  For simplicity, suppose that one of them is
instantaneously at rest, while the other is moving.
During a short time interval $\delta t$, the moving
segment will trace out a thin ribbon of width
$|\dot{\bf x}|\delta t$.  Intercommuting will occur if the
other segment intersects the ribbon.  So long as
$\delta t$ is small compared to the small-scale structure,
the probability of this is clearly proportional to
the lengths of the segments and is in no way reduced
by the fact that they are kinky.

What is true of kinky strings is that where one
intersection occurs, others are likely in the vicinity.
This means that in the neighbourhood of an
intercommuting loop formation is likely to occur.

\subsection{Effect of intercommuting}

We now turn to the effect of long-string intercommuting
on the probability distribution $p[{\bf r}(l)]$.

Clearly there is a negative term in
$(\partial (Lp)/\partial t)_{\rm l.s.i.}$ due to the
destruction of segments by intercommuting, equal to
\begin{equation}
-\chi {l\over\xi^2} Lp[{\bf r}(l)].
\label{lsineg}
\end{equation}

In this case, the corresponding positive term is rather
more complicated.  For each segment destroyed, a new
segment of equal length $l$ is created.  However, its
extension ${\bf r}$  is the sum of the extensions
${\bf r}_1$ and ${\bf r}_2$ of the two, generally speaking
uncorrelated, segments thus brought together.

The number of segments created is exactly equal to the
number destroyed.  The probability that one of these,
chosen at random, is composed of lengths $l_1$ and
$l-l_1$, within an interval $dl_1$, is $dl_1/l$.
Then the probability that the corresponding extensions
are ${\bf r}_1$ and ${\bf r}_2$, within intervals
$d^3{\bf r}_1$ and $d^3{\bf r}_2$ (see
Fig.\ \ref{fig3}), is
\begin{equation}
p[{\bf r}_1(l_1)] d^3{\bf r}_1\,p[{\bf r}_2(l-l_1)]d^3{\bf r}_2.
\end{equation}
To find the probability that the total extension is
{\bf r}, we have to set ${\bf r}_2 = {\bf r} - {\bf r}_1$ and
integrate over ${\bf r}_1$.  Thus the positive term in
$(\partial (Lp)/\partial t)_{\rm l.s.i.}$ is
\begin{equation}
+ {\chi l L\over \xi^2} \int_0^l {dl_1\over l}
\int d^3{\bf r}_1 \, p[{\bf r}_1(l_1)]\,
p[({\bf r}-{\bf r}_1)(l-l_1)].
\label{lsipos}
\end{equation}
In contrast to the case of loop formation, here the
assumption that the two probabilities are independent
should be a good one, because the two segments involved
generally come from regions that are far apart along the
string.

In this case, we evidently have
\begin{equation}
\left(\partial L\over\partial t\right)_{\rm l.s.i.} = 0.
\end{equation}
Hence, putting (\ref{lsineg}) and (\ref{lsipos})
together, we obtain
\begin{equation}
\left(\partial p\over\partial t\right)_{\rm l.s.i.} =
{\chi\over\xi^2} \left\{\int_0^l dl_1
\int d^3{\bf r}_1\, p[{\bf r}_1(l_1)]\,
p[({\bf r}-{\bf r}_1)(l-l_1)] - lp[{\bf r}(l)]\right\}.
\label{pdlsi}
\end{equation}

It is again straightforward to find an expression
for the rate of change of $K$.  When we multiply by
${\bf r}^2$ and integrate, in the first term it is
best to go back to using ${\bf r}_1$ and ${\bf r}_2$ as
integration variables.  Since there is assumed to
be no correlation the mean value of
${\bf r}_1{\cdot}{\bf r}_2$ vanishes.  Thus we find
 \begin{equation}
\left(\partial K\over\partial t\right)_{\rm l.s.i.} =
{\chi\over\xi^2} \left\{\int_0^l dl_1
[K(l_1)+K(l-l_1)] - lK(l)\right\}.
\label{Kdlsi1}
 \end{equation}

Note that in general we expect
 \begin{equation}
K(l) > K(l_1) + K(l-l_1),
 \end{equation}
so the effect of long-string intercommuting is to
{\it reduce\/} the value of $K$ (unless $K$ is a linear
function of $l$, in which case it is unchanged).

By symmetry, we can simplify the expression
(\ref{Kdlsi1}) slightly:
 \begin{equation}
\left(\partial K\over\partial t\right)_{\rm l.s.i.} =
{\chi\over\xi^2} \left\{2\int_0^l dl_1
K(l_1) - lK(l)\right\}.
\label{Kdlsi}
 \end{equation}

 From this, we can find the rates of change of the
various length scales.  Intercommuting has no effect on
$L$, or, therefore, on $\xi$.

 The length scale $\bar\xi$ is defined by
(\ref{Kbigl}).  To find its rate of change we
differentiate (\ref{Kdlsi}) and then allow $l$ to
approach infinity:
 \begin{equation}
2{d\bar\xi\over dt} =
\lim_{l\to\infty}
{\partial K'(l)\over\partial t}.
\label{xbd}
 \end{equation}
This yields
 \begin{equation}
{\dot{\bar\xi}_{\rm l.s.i.}\over\bar\xi} =
- {\chi\over w}{\bar\xi\over\xi^2}.
\label{xbdlsi}
 \end{equation}

Similarly,  from (\ref{Ksmall}) we find
 \begin{equation}
{2\dot\zeta\over\zeta^2} =\lim_{l\to0}
{\partial K'''(l)\over\partial t}.
 \label{zed}
 \end{equation}
Of course, for this to be consistent, we also have to
verify that the rates of change of $K$, $K'$ and $K''$
all vanish in the limit.  This is easy to do.  Thus we
find
 \begin{equation}
{\dot{\zeta}_{\rm l.s.i.}\over\zeta} =
- {\chi\zeta\over\xi^2}.
\label{zedlsi}
 \end{equation}

\subsection{Effect of loop formation}

Next we review and revise the derivation of the
probability of loop formation.    This is
perhaps the most problematic aspect of our analysis; we
shall approach it in several stages.

Consider a segment of string of length
$l$ and extension ${\bf r}$.  We want to evaluate the
probability that this segment will form a loop during a
short time interval.  To be specific, let $\Theta({\bf r},l)
\,dl\,dy\,dt$ be the probability that a loop will form
in the time interval $dt$, of length between $l$ and
$l+dl$, and with starting point within the small
interval $dy$.  This is the function we aim to
estimate.

 From $\Theta$, we can compute rates of change due to
loop formation of all the quantities we need. In
particular, the probability that any particular point
on the string will be incorporated into a loop within
the time interval $dt$ is clearly $\lambda dt$, where
 \begin{equation}
\lambda = \int_0^\infty dl\,l\Lambda(l),
\label{lamdef}
 \end{equation}
and
 \begin{equation}
\Lambda(l) = \int d^3{\bf r}
\,p[{\bf r}(l)]\Theta({\bf r},l).
\label{Lamdef}
 \end{equation}

The parameter $\lambda$ determines the overall rate of loss
of length to loop formation.
Consider a very long section of string, of total length
$L$.  Then the expected rate of change of $L$ is
 \begin{equation}
\left(\partial L\over\partial t\right)_{\rm loops}
= - \lambda L.
\label{Ldllf}
 \end{equation}

The  dependence of $\Theta$ on ${\bf r}$ is unknown
and may well be complicated. However, consistent with our
Gaussian  approximation, it seems reasonable to assume
that, except for very small values of $l$, this is also
Gaussian, parametrized by a variance function $Q(l)$.
Specifically, we assume that
 \begin{equation}
\Theta({\bf r},l) \approx \Lambda(l)\left(K(l)+Q(l)\over
Q(l)\right)^{3/2}
e^{-3{\bf r}^2/2Q(l)}.
\label{Qdef}
 \end{equation}

We now turn to the expression for the effect of
loop formation on the evolution of $p[{\bf r}(l)]$.
Let us consider the expression for
$(\partial(Lp)/\partial t)_{\rm loops}$.  It consists of a negative
term representing the number of segments lost to loop
formation and a corresponding positive one representing
the number of new segments created by the process.

The chosen segment of length $l$ and extension
${\bf r}$ will disappear if a loop is formed
within it, overlapping either end, or enclosing it
entirely.  Let us ask for the probability that this
happens within a short time interval $dt$ due to the
formation of a loop of length $l_1$, within a range
$dl_1$.

We denote the starting point of the loop segment
relative to that of our chosen segment of length
$l$ by $y_0$ (see Fig.\ \ref{fig4}).  Allowing for
all possible overlaps, the range of possible starting
points is
 \begin{equation}
-l_1 < y_0 < l.
 \end{equation}
[The total number of starting points, with discretization
scale $\delta$, is
$(l+l_1)/\delta$.]
The probability that a segment of length $l_1$
starting from one of these points has extension
${\bf r}_1$ within a volume $\delta^3$ is
$p[{\bf r}_1(l_1)|{\bf r}(l),y_0]\delta^3$,
where the symbol $p[\dots|\dots]$ denotes a
probability conditional on both the value of
${\bf r}(l)$ and on the position of the starting
point $y_0$.  Rather than using this conditional
probability, it is more convenient to
work with the corresponding joint probability, using
the identity
 \begin{equation}
p[{\bf r}_1(l_1),{\bf r}(l)|y_0] =
p[{\bf r}_1(l_1)|{\bf r}(l);y_0]\,p[{\bf r}(l)].
 \end{equation}
However, the joint probability is still conditional on
the position $y_0$ of the starting point.

The probability that within a time interval $dt$ a loop
is formed of length between $l_1$ and $l_1+dl_1$,
starting between $y_0$ and $y_0+dy_0$ is, according to
the discussion of Section V, $\Theta({\bf r}_1,l_1)\,
dl_1\,dy_0\,dt$.

Thus we find for the negative term  the expression
 \begin{equation}
-L\int_0^\infty dl_1\, \int d^3{\bf r}_1 \,
\Theta({\bf r}_1,l_1)
\int_{-l_1}^l dy_0\,
 p[{\bf r}_1(l_1),{\bf r}(l)|y_0].
\label{Lpdllneg}
 \end{equation}

In evaluating the corresponding positive term, we
have to consider excision of a loop entirely within
a segment of length $l+l_1$.  Thus it is
 \begin{equation}
L\int_0^\infty dl_1\, \int d^3{\bf r}_1 \,
\Theta({\bf r}_1,l_1)
\int_0^l dy_0 \,
 p[{\bf r}_1(l_1),({\bf r}+
{\bf r}_1)(l+l_1)|y_0].
\label{Lpdllpos}
 \end{equation}

Putting (\ref{Lpdllneg}) and (\ref{Lpdllpos})
together, we have
 \begin{eqnarray}
\left(\partial(Lp)\over\partial t\right)_{\rm loops}
&=&L\int_0^\infty dl_1\, \int d^3{\bf r}_1 \,
\Theta({\bf r}_1,l_1)
\biggl\{\int_0^l dy_0
\, p[{\bf r}_1(l_1),({\bf r}+
{\bf r}_1)(l+l_1)|y_0]
\nonumber\\
&-&
\int_{-l_1}^l dy_0
\, p[{\bf r}_1(l_1),{\bf r}(l)|y_0]
\biggr\}.
\label{Lpdll}
 \end{eqnarray}

It is straightforward
to perform the integration over
${\bf r}$  to find $(\partial L/\partial t)_{\rm loops}$,
because clearly
 \begin{equation}
\int d^3{\bf r} \, p[{\bf r}_1(l_1),
{\bf r}(l)|y_0] =
p[{\bf r}_1(l_1)],
\label{ldll}
 \end{equation}
independent of $y_0$.  This of course reproduces
(\ref{Ldllf}) with $\lambda$ given by (\ref{lamdef}).

Combining (\ref{Lpdll}) and (\ref{Ldllf}) we obtain
 \begin{eqnarray}
\left(\partial p\over\partial t\right)_{\rm loops}
&=&\int_0^\infty dl_1\, \int d^3{\bf r}_1 \,
\Theta({\bf r}_1,l_1)
\biggl\{\int_0^l dy_0
\, p[{\bf r}_1(l_1),({\bf r}+
{\bf r}_1)(l+l_1)|y_0]
\nonumber\\
&-&
\int_{-l_1}^l dy_0
\, p[{\bf r}_1(l_1),{\bf r}(l)|y_0]
+ l_1 p[{\bf r}_1(l_1)]p[{\bf r}(l)]
\biggr\}.
\label{pdll}
 \end{eqnarray}

\subsection{Probability of loop formation}

Now let us turn to the calculation of
$\Theta({\bf r},l)$, or
equivalently $\Lambda(l)$ and $Q(l)$. Our segment
of length $l$ will form a loop during a  short time
interval $dt$ if the corresponding total  extension
${\bf r}_{\rm tot}$, given by (\ref{rtot}), vanishes
at some instant during that interval.  In other words,
the corresponding left- and right-moving segments,
each of length $l$, must have the same extension,
${\bf r}$.

To simplify the counting, let us imagine that
space-time is partitioned into cells each of volume
$\delta^4$, and moreover that the string is partitioned
into small segments of length $\delta$.

The probability that a left-moving segment of length
$l$ has an extension ${\bf r}$ within a $\delta^3$
volume labelled $j$ is
\begin{equation}
p_j = p[{\bf r}(l)] \delta^3.
\end{equation}
The probability that a loop is formed is essentially
the probability that the corresponding right-moving
segment also has the same extension.  Assuming for the
moment that the probabilities are uncorrelated (which as
we shall see is by no means true), this probability
is also $p_j$.  We have to multiply it by factors
relating to $dl$, $dt$, etc.   The
number of length steps within a given range $dl$ is
$dl/\delta$.  The number of time steps within $dt$ is
$2dt/\delta$.  (The factor of 2 arises, as explained in KC
\cite{KC}, because the segments are moving with the speed
of light in opposite directions, so that each encounters
a new segment after a time $\delta/2$.)  The number of
starting points on $dy$ is $dy/\delta$.

However, we should not merely multiply these
factors.  Particularly for small loops, the angles
between the various vectors are small.

A typical configuration is shown in Fig.\ \ref{fig5}.
The question is, by how much can we vary the length
$l$, the starting point $y$ and the time $t$ without
moving the difference in extensions of the two segments
out of the $\delta^3$ volume?  This is essentially the
same calculation that we did earlier in estimating the
parameter $\chi$ that determines the rate of
long-string intercommuting.  The volume swept out by
the total extension ${\bf r}_{\rm tot}$ when the parameters
vary over ranges $dl$, $dy$ and $dt$ is

 \begin{equation}
|{\bf x}_1'\wedge{\bf x}_2'{\cdot}
(\dot{\bf x}_1 - \dot{\bf x}_2)|\,dl\,dy\,dt.
\label{sctrpr2}
\end{equation}
Thus we should include in
our expression for the loop-formation
probability a factor

 \begin{equation}
\Delta({\bf r},l) = \case{1}/{4}
\langle|{\bf p}_1\wedge{\bf p}_2{\cdot}
({\bf q}_1 - {\bf q}_2) + ({\bf p}_1 - {\bf p}_2)
{\cdot}{\bf q}_1\wedge{\bf q}_2|\rangle.
\label{deltadef}
 \end{equation}

The number of cells within
the volume $d^3{\bf r}$ is $d^3{\bf r}/\delta^3$.  Hence
our required probability is
 \begin{equation}
p[{\bf r}(l)]\Theta({\bf r},l)\,d^3{\bf r}\,dl\,dy\,dt
= p_j^2\,{d^3{\bf r}\over\delta^3}
{dl\over\delta}{2dt\over\delta}{dy\over\delta}
\Delta({\bf r},l),
 \end{equation}
{\it i.e.},
 \begin{equation}
\Theta({\bf r},l) = 2p[{\bf r}(l)]\Delta({\bf r},l).
\label{Th1}
 \end{equation}
Note that the factors of $\delta$ cancel, as they must.

Inclusion of the factor $\Delta$ eliminates the problem
of multiple counting of loops noted in KC \cite{KC}.
This was avoided in CKA \cite{CKA} by imposing a
small-scale cutoff and treating the contribution of
small loops separately.  However, that procedure
introduced additional problems associated with the
choice of the cutoff.  It is better to treat loop
production within a unified framework.

So long as the loops are reasonably large, it is
reasonable to assume that the unit vectors involved are
independently randomly distributed on the unit sphere.
Then we can replace $\Delta$ by its average,
$\overline\Delta = \chi \approx 0.2$.

For very small loops we expect $\Delta$ to be small.
Consider for example the model in which the strings are
composed of straight segments joining randomly
distributed kinks.  For very short loops one would
expect to find only a single kink on each of the left-
and right-moving segments forming the loop.  However, it
is easy to see that in this case the scalar triple
product (\ref{sctrpr2}) vanishes identically, because
the three vectors are coplanar.  This is because a
triangular loop is necessarily planar.
The point is that the loop-formation condition in this
situation is satisfied only on a set of configurations
of measure zero.

This is of course an idealized model, but even in  a
more realistic model we should expect that the $\Delta$
factor would be very small for loops with only a single
pair of kinks.

It follows from this argument that $\Delta$ should
vanish rapidly as $l\to0$. In the simple model, the
leading contribution would come from loops with at
least three kinks all together.  If the kinks are
randomly distributed, with a separation of order
$\zeta$, then the number of kinks on a segment of
length $l$ is Poisson distributed, with mean $\approx
l/\zeta$, so for three or more kinks we expect a
factor of at least $(l/\zeta)^3$.

Also for small loops, the ${\bf r}$-dependence is likely
to be important.  For values of $|{\bf r}|$ close to $l$,
the segments forming the loop must be nearly
straight with a high degree of correlation between the
${\bf p}$ vectors at the two ends, and also between the
${\bf p}$ and ${\bf q}$ vectors.  In that case the scalar
triple product will be small.  On the other hand
smaller values of ${\bf r}$  imply large-angle kinks and
correspondingly less correlation between the unit
vectors.  So we expect $\Delta$ to decrease with
increasing $|{\bf r}|$.

Except for very small loops, where $\Delta$ is in
any case small, ${\bf r}^2$ is generally much less than
$l^2$.  Thus it would seem reasonable to assume,
consistent with our earlier assumptions, that it has
the form of a Gaussian,
 \begin{equation}
\Delta({\bf r},l) \approx \Delta_0 e^{-a{\bf r}^2/l^2},
\label{delta0def}
\end{equation}
where $a$ is a constant of order unity.  The overall
factor $\Delta_0$ approaches $\chi$ at large $l$ and
vanishes at least like $(l/\zeta)^3$ as $l \to 0$.  The
Gaussian factor yields a contribution to $1/Q(l)$ of
magnitude $2a/3l^2$.

The formula (\ref{Th1}) is still not accurate, for
several reasons, but  particularly because it neglects
all  correlations between the left- and
right-moving strings.  The rate of change of $L$ due to
loop formation is to be computed using (\ref{Ldllf})
and (\ref{lamdef}).
As it stands, this expression would
give problems at both ends of the range of integration
over $l$.  At the upper
end, we have to consider the probability of
reconnection; we aim to include only loops that do not
reconnect.  At the lower end, the integral
would diverge, because of the
neglect of a very significant angular correlation effect.

We deal first with reconnection.
For a loop of size $l$, the probability of reconnection
within a short time interval $dt$ is
$\chi l dt/\xi^2$.  The probability that the
loop will survive reconnection to a much later time is
therefore
\begin{equation}
\exp\left(-\chi\int_t^\infty dt'\,
{l\over\xi(t')^2}\right).
\label{recon1}
\end{equation}
If we assume that over the relevant period $\xi$ at least
approximately scales, {\it i.e.}, $\xi(t) \propto t$, then
the integral yields
\begin{equation}
\exp\left(-{\chi lt\over\xi(t)^2}\right).
\label{recon}
\end{equation}
(It is not necessary here to allow for the variation
of $l$ with time, due to gravitational radiation,
which occurs over a very much longer time scale.)
To allow for the probability of reconnection, we
should replace (\ref{Th1}) by
\begin{equation}
\Theta({\bf r},l) = 2p[{\bf r}(l)]
e^{-\chi lt/\xi^2}\Delta({\bf r},l).
\label{Th2}
\end{equation}
The exponential provides an effective upper cutoff
in (\ref{lamdef}) at a scale
of order $\xi^2/t$.  Above that scale, loops are
rather likely to reconnect; below it, they mostly
survive.

It should be noted that the assumption of approximate
scaling is a very weak one.  Even if $\xi$ does not
exactly scale, the dominant contribution to the integral
(\ref{recon1}) will come from close to the lower limit
so (\ref{recon}) will change at most by a factor
of order unity.

Even with the
modifications described, the  expression
(\ref{Th2}) is still not entirely correct, because
it neglects any possible correlation between the
left- and right-moving segments.  This turns out to be
the most important effect of all.  These correlations
form the subject of the next Section.

There is a potentially important correction yet to be
included.  Some of the long-string intercommuting
events create large loops.  (Small loops tend to be
created by a rather different process, very dependent on
the small-scale structure on the strings; there is less
likelihood of confusion.)  Unless these loop-forming
crossings can somehow be
excluded, there is a possibility of double counting.
It is not immediately clear how serious the problem is
likely to be, although one might guess that it affects
positive and negative contributions in a similar
way, so that the overall effect may not be very large.

\section{CORRELATIONS BETWEEN LEFT- AND RIGHT-MOVERS}

  Taking account of the correlations between
left- and right-moving segments,  we
should really write
  \begin{equation}
\Theta({\bf r},l)p[{\bf r}(l)] = 2p[{\bf r}(l);{\bf r}(l)]
e^{-\chi lt/\xi^2}
\Delta({\bf r},l),
\label{Th4}
 \end{equation}
where $p[{\bf r}(l);{\bf r}'(l)]
\,d^3{\bf r}\,d^3{\bf r}'$ is the joint probability that the
corresponding left- and right-moving segments have
extensions ${\bf r}$ and ${\bf r}'$ respectively.

 One approach might be to extend the
Gaussian Ansatz, representing the joint probability
distribution by a six-dimensional Gaussian with an
appropriate covariance function
$\overline{{\bf r}\cdot{\bf r}'}$.
However, this is not a good representation of
$p[{\bf r}(l);{\bf r}'(l)]$.  The covariance is actually quite
small (and negative), suggesting that the correlation
effect yields merely a small reduction in the
loop-formation probability.  But this is quite false:
the joint probability distribution is in fact very
sharply reduced in the forward
$({\bf r}={\bf r}')$ direction.

We begin by considering the correlation of the
individual ${\bf p}$ and ${\bf q}$ vectors.
There are two  quite
separate processes that generate such correlations.  In
CKA \cite{CKA} we considered only the effect of
stretching, but in fact loop production also plays an
important role, and indeed intercommuting cannot be
ignored.

Consider a particular ${\bf p}$ segment, ${\bf p}(u_0)$,
and the approaching
${\bf q}$ segments, in  particular a segment
${\bf q}(v_0)$ which encounters it at time $t_0$.  (See
Fig.\ \ref{fig6}.)   We wish
to estimate how the angular probability distribution
of ${\bf q}$ relative to the direction of ${\bf p}$ changes as
the vectors approach one another. It is again
convenient to use the path-length variable $y$ defined
in (\ref{ydef}).  Let us define $\Phi(y,z)$, so that the
probability that  ${\bf p}(y_0){\cdot}{\bf q}(y_1) = z$, within
the range $dz$, is
 \begin{equation}
\Phi(y,z) {dz\over2},
 \end{equation}
where $y = y_0-y_1$.
With this normalization,
the initial  condition for $\Phi$ at large
$y_0-y_1$ is
 \begin{equation}
\Phi(\infty,z) = 1,
 \end{equation}
representing a completely random distribution
when the segments are far apart.

Let us now seek to write down an evolution
equation for $\Phi$, of the form
 \begin{equation}
{\partial\Phi\over\partial y} =
\left({\partial\Phi\over\partial y}\right)_{\rm str.}
+ \left({\partial\Phi\over\partial y}\right)_{\rm l.s.i.}
+ \left({\partial\Phi\over\partial y}\right)_{\rm loops}.
\label{dphi}
 \end{equation}
We are assuming here that the process of
establishing an angular correlation as the
vectors ${\bf p}$ and ${\bf q}$
approach takes a relatively short time, compared
to the time scales for evolution, so that $\Phi$
may be regarded as a function only of $z$ and of the
path-length $y$ between them, not explicitly of
the time.  This is reasonable because the angular
correlation sets in only when the vectors are
already quite close.  For similar reasons, we are
justified in neglecting the effect of gravitational
radiation which is small on the scales of
interest  here.

We also introduce the corresponding angular
probability distribution for the ${\bf p}$ vector
that encounters ${\bf q}$ at $y_1$: the probability
that ${\bf p}(y_0){\cdot}{\bf p}(y_1) = z$ within the range
$dz$ is
 \begin{equation}
\Psi(y,z)\,{dz\over2}.
 \end{equation}

Let us recall that
 \begin{equation}
\bar z_\Psi \equiv \int_{-1}^1 {dz\over2}\,
z\Psi(y,z) = f(y),
 \end{equation}
where $f$ is the function defined in (\ref{fdef}).
(When it is necessary to distinguish, we denote the
mean value of $z$ with respect to the
distribution $\Phi$ by $\bar z_\Phi$ and that with respect
to $\Psi$ by $\bar z_\Psi$.)

\subsection{The exponential Ansatz}

To reduce the problem to
manageable proportions, we make a simplifying
assumption concerning the angular distribution
function $\Psi$, analogous to the
Gaussian Ansatz, namely that  (except when $y$ is
very small)  it takes the form of an exponential:
 \begin{equation}
\Psi(y,z) = {k\over\sinh k}
e^{kz},
\label{psiexp}
 \end{equation}
where $k$ is a function of $ y$.  The relation
between $k$ and $f$ is
 \begin{equation}
\bar z_\Psi = f( y) = \coth k
- {1\over k}.
 \end{equation}
 (As we shall see the Ansatz breaks down for
very small values of $y$.)

It will be useful to examine the limiting cases
of large and small values of $ y$.
First, when
$ y$ is large, $f\ll1$ and consequently
also $k\ll1$.  In that case we have
  \begin{equation}
f \approx {k\over3}
- {k^3\over45}
\qquad ( y\;{\rm large}).
\label{fbigy}
  \end{equation}
In most cases, therefore, a linear approximation
$k \approx 3f$ will be adequate.  Neglecting
$k^2$, we may write
  \begin{equation}
\Psi \approx 1 + kz \approx 1 + 3fz
\qquad ( y\;{\rm large}).
\label{psibigy}
  \end{equation}

In the opposite limit, where $ y$ is small (but not
small enough to render the Ansatz invalid), we have
$1-f \ll 1$ and hence $k \gg 1$.  In this case, the
leading approximation is
  \begin{equation}
k \approx {1\over 1-f} \approx {\zeta\over y}
\qquad ( y\;{\rm small}).
  \end{equation}
Here we may write
\begin{equation}
\Psi \approx 2k e^{-k(1-z)}
\qquad ( y\;{\rm small}).
\label{psismy}
\end{equation}
The distribution becomes concentrated near $z = 1$
within a range of order $1/k$.

It is interesting to note that in the intermediate
region, both approximations are in fact reasonably
good.  For example for $f = {1\over2}$, the
leading large-$ y$ and small-$ y$
approximations give
respectively $k = {3\over2}$ and $k=2$; the
correct answer is $k=1.8$.

When the exponential Ansatz is valid, all the moments
of the distribution are of course determined by the
expectation value.  In particular,
 \begin{eqnarray}
\big(\overline{z^2}\big)_\Psi
&=& 1 - {2\coth k\over k}
+ {2\over k^2}
\nonumber\\
&\approx& {1\over3} + {2 k^2\over45}
\approx {1\over3} + {2 \bar z_\Psi^2\over5}
\qquad ( y\;{\rm large}).
 \end{eqnarray}
Indeed, so long as $k\ll1$ it is a good approximation
to set $\big(\overline{z^2}
\big)_\Psi \approx {1\over3}$.

The exponential Ansatz can also be applied to
the ${\bf q}$ distribution $\Phi$, in the modified form

 \begin{equation}
\Phi( y,z) = {b\over\sinh b}e^{-bz}.
\label{phidef}
 \end{equation}
(Since the directions of ${\bf p}$ and
${\bf q}$ vectors are
{\it anti}-correlated, with this definition $b$
is positive.)    However, as we shall see,
the Ansatz ceases to be a
good approximation for very small values
of $y$.  When the approximation is valid,
the evolution equation for $\Phi$
effectively reduces to an equation for $b$
or equivalently $\bar z_\Phi$.

\subsection{Equation for $\Phi$}

Now let us consider the ensemble of approaching ${\bf q}$
vectors, in particular those that at a given time
$t$ fall within a small interval $\delta y_1$ at
$y_1$.

In the time interval $dt$, these vectors
either move closer to $y_0$ or else are eliminated
by being incorporated into a loop.  The expected
distance by which they move closer is
\begin{equation}
dy = -(2+\lambda y)dt,
\label{dy1}
\end{equation}
with $y = y_0-y_1$.  Here the 2 is the normal
velocity of approach and the extra term $\lambda y dt$
represents  the expected loss of length between $y_1$
and $y_0$ due to loop formation.  Integrating this
relation, we find
\begin{equation}
y  = {2\over\lambda}\left(
e^{\lambda(t_0-t)} - 1\right),
\end{equation}
 where $t_0$ is the time at which the segments
coincide.  Note that the distance $y$ becomes
exponentially large for time differences larger
than $1/\lambda$.  Loop formation effectively
shuffles the string segments on this time scale,
providing a very effective long-distance cut-off.

During the time interval $dt$, the small interval
$\delta y_1$ on the string effectively becomes  a
little shorter, because some members of the ensemble are
eliminated by incorporation into loops.  In fact,
$d\delta y_1/dt = - \lambda\delta y_1$, so
$\delta y_1$ is replaced by
$\delta y_1 - \lambda dt\,\delta y_1$.
The angular distribution
of the remaining vectors changes from
$\Phi(y,z)$ to $\Phi(y+dy,z)$, with $dy$
given by (\ref{dy1}).  To find the difference
between these, we need to know the angular
distribution, ${\rm X}( y,z)$ say, of the vectors
that have been
eliminated.  Here ${\rm X}(y_0-y_1,z) {dz\over2}$ is
the probability that a vector ${\bf q}$ which
is incorporated into a loop within a short
time interval has
${\bf p}(y_0){\cdot}{\bf q}(y_1) = z$ within
the range $dz$.  If we can estimate ${\rm X}$, then
we can obtain a differential equation for $\Phi$.
In fact (ignoring stretching and intercommuting
contributions),
 \begin{equation}
\delta y_1\,\Phi(y,z)
= (\delta y_1 - \lambda dt\,\delta y_1)\,\Phi(y+dy,z)
+ \lambda dt\,\delta y_1\,{\rm X}(y,z),
\label{dPhillf}
 \end{equation}
or equivalently
 \begin{equation}
(2+\lambda y)\left(
\partial\Phi\over\partial y\right)_{\rm loops}
= - \lambda\Phi + \lambda{\rm X}.
\label{dPhidyllf}
 \end{equation}
  Note that this is consistent with maintenance of
the normalization condition $\int{dz\over2}\Phi = 1$
provided that ${\rm X}$ is also normalized.

The key observation is that since loops are formed by
matching segments of left- and right-moving string,
the angular distributions of excised
${\bf p}$ and ${\bf q}$
segments should be identical.  Clearly, the probability of
obtaining a matching pair must depend on the probability
distributions of both ${\bf p}$ and ${\bf q}$.
It seems reasonable to assume that ${\rm X}$ is
proportional to the {\it product} $\Psi\Phi$.
Normalization then requires
 \begin{equation}
{\rm X}(y,z) = N(y)\Psi(y,z)\Phi(y,z),
 \end{equation}
where
 \begin{equation}
N^{-1}(y) = \int_{-1}^1 {dz\over2}
\Psi(y,z)\Phi(y,z).
\label{Normdef}
 \end{equation}
Thus we find
 \begin{equation}
(2+\lambda y)\left(\partial\Phi\over
\partial y\right)_{\rm loops} = - \lambda\Phi
+ \lambda N\Psi\Phi.
 \end{equation}

So far we have ignored both intercommuting and stretching,
both of which will also have an effect on the angular
distribution $\Phi$.  Consider first the effect of
intercommuting.  The probability that an intercommuting
event occurs between $y_1$ and $y_0$ within the time
interval $dt$ is $\chi y dt/\xi^2$.  If it does occur
the relevant members of the ensemble of
${\bf q}$ vectors
are deleted and replaced by new vectors drawn from an
essentially independent random distribution.  In other
words, $\Phi(y,z)$ on the  left-hand side of
(\ref{dPhillf}) is replaced by
 \begin{equation}
\Phi( y,z)\left(1 -
{\chi y dt\over\xi^2}\right)
+ {\chi y dt\over\xi^2}.
 \end{equation}
Thus the effect of intercommuting is described by
 \begin{equation}
(2+\lambda y)\left(\partial \Phi
\over\partial y\right)_{\rm l.s.i.}
= -{\chi y\over\xi^2}(1-\Phi).
 \end{equation}

Finally, let us consider the effect of stretching.
The equation
of motion (\ref{eqmotq}) for ${\bf q}$ may be
written
 \begin{equation}
{\partial{\bf q}\over\partial t} =
- H({\bf p}_1
- {\bf q}{\bf q}{\cdot}{\bf p}_1).
\label{eqmotqy}
 \end{equation}
where  ${\bf p}_1$ denotes the
vector ${\bf p}(y_1)={\bf p}(u_1,v_0)$ (see
Fig.\ \ref{fig6}).   Strictly speaking, the Hubble
parameter  $H$ is a variable quantity here, but,
since the correlations extend over distances which are
small compared to the horizon distance,
it should be a good approximation to treat
it as a constant.

We now have to average this result over the
angular distribution of ${\bf p}(y_1)$.
In principle, this vector is correlated with both
${\bf p}(y_0)$ and ${\bf q}(y_1)$.
However, any contribution proportional to
${\bf q}(y_1)$ cancels out in (\ref{eqmotqy}).
Thus it is reasonable to consider only its correlation
with ${\bf p}(y_0)$, and to replace
${\bf p}(y_1)$ by $\bar z_\Psi{\bf p}(y_0)$.

There is an exactly similar expression for the rate
of change of ${\bf p}(y_0)$, which yields an identical
contribution.  Taking account of both we thus find
 \begin{equation}
(2+\lambda y)\left(
{\partial\Phi\over\partial y}\right)_{\rm str.}
= {\partial\over\partial z}\left(
{\partial z\over\partial t} \Phi\right)
= -2H\bar z_\Psi
{\partial\over\partial z}\left[
(1-z^2) \Phi\right].
 \end{equation}

Bringing all three contributions together,
we may write the equation
for $\Phi$ as
 \begin{equation}
(2+\lambda  y){\partial\Phi\over\partial y}
= -2H\bar z_\Psi
{\partial\over\partial z}\left[
(1-z^2) \Phi\right]
- {\chi y\over\xi^2}(1-\Phi)
- \lambda\Phi + \lambda N\Psi\Phi.
\label{dPhitot}
 \end{equation}
When the exponential Ansatz is valid, all we need
is the equation
for the rate of change of $\bar z_\Phi$, namely

 \begin{equation}
(2+\lambda y){\partial\bar z_\Phi\over\partial y}
=  2H\bar z_\Psi\big[1-
\big(\overline{z^2}\big)_\Phi
\big] + {\chi y\over\xi^2}\bar z_\Phi
- \lambda\bar z_\Phi + \lambda\bar z_{\rm X},
\label{zbard}
 \end{equation}
 where $\bar z_{\rm X}$ is the mean with respect to
the  distribution ${\rm X} = N\Psi\Phi$.

\subsection{Equation for $\Psi$}

Before trying to solve the equation for $\Phi$,
some parenthetical remarks about the possibility of
deriving a similar equation for $\Psi$ may be
in order.

The effects of stretching and intercommuting on
$\Psi$ are very similar, and we have assumed that
the distribution ${\rm X}$ of excised segments
is the same for both.  So we should be able to write
down a very similar equation for the evolution of
$\Psi$.  However, there is a very important
difference, concerning the rate at which one
${\bf p}$ segment approaches another.
The most obvious difference is that
the distance $y = y_0 - y_1$ between the two ${\bf p}$
vectors decreases {\it only\/} because of loop formation.
In other words the term 2 in the factor on the left
of (\ref{dPhidyllf}) is absent in the corresponding
equation for $\Psi$.

But there is a further point: it is
no longer reasonable to assume that $dy/dt$ is
independent of $z$.  In fact, if it were, the
string would never develop the long-range directional
correlation that it does.
The formation of loops (except for the very smallest)
depends strongly on the large-scale
configuration of the strings.  If the
left- and right-moving sections are relatively straight,
the large-loop formation probability is low, because
the values of ${\bf r}$ for given $l$ are large.  On
the other hand, if the strings are curled up tightly,
typical values of ${\bf r}$ are
small and large-loop formation
becomes highly probable.  In the former case, positive
values of ${\bf p}(y_0){\cdot}{\bf p}(y_1)$ are
clearly favoured.  Conversely, if
${\bf p}(y_0){\cdot}{\bf p}(y_1)$ is positive, the
expected value of $-dy/dt$
will be smaller than
if it is negative.  Roughly speaking, we may expect
 \begin{equation}
{dy\over dt} \approx
- [\lambda y - (z - \bar z_\Psi)v(y)],
 \end{equation}
where $v(y)$ is a function that could in principle
be determined from the later discussion of the
detailed effects of loop formation.
There is no doubt a similar effect even for the
cross-correlation between ${\bf p}$ and ${\bf q}$,
but in that case it seems likely to be negligibly
small.

It follows that for $\Psi$, in place of (\ref{dPhitot}),
we would have
 \begin{equation}
{\partial\over\partial y}
\Big(\big[\lambda y - (z -
\bar z_\Psi)v(y)\big]\Psi\Big)
= - 2H\bar z_\Phi
{\partial\over\partial z}\left[
(1-z^2) \Psi\right]
- {\chi y\over\xi^2}(1-\Psi)
+ \lambda N\Phi\Psi.
\label{dPsitot}
 \end{equation}
 Similarly, the analogue of (\ref{zbard}) is
 \begin{equation}
{\partial\over\partial y}
\Big(\lambda y\bar z_\Psi -
\big[\big(\overline{z^2}\big)_\Psi
- \bar z_\Psi^2\big]v(y)\Big)
= 2H\bar z_\Phi\big[1-
\big(\overline{z^2}\big)_\Psi
\big] +{\chi y\over\xi^2}\bar z_\Psi
+ \lambda\bar z_{\rm X}.
\label{zbarPsid}
 \end{equation}

\subsection{Solution of equation for $\Phi$}

Having set up the equation (\ref{dPhitot}) for
$\Phi$ we now set
about solving it.  For the moment at least, we
shall treat $\Psi$ as given, via the exponential
Ansatz, in terms of the correlation function
$\bar z_\Psi = f(y)$.

Consider first the region of large $y$ where the linear
approximation (\ref{psibigy}) for $\Psi$ and $\Phi$
should be valid. In this case, we need
consider only the evolution equation for $\bar z_\Phi$.
So long as $\bar z_\Phi$ remains
small, we can also use the approximation
$(\overline{z^2})_\Phi \approx {1\over3}$.  Note also that
under these conditions, the distribution ${\rm X}$
is also linear, with
 \begin{equation}
\bar z_{\rm X} = \bar z_\Psi + \bar z_\Phi.
\label{zbarX}
 \end{equation}

Thus we find
 \begin{equation}
(2+\lambda y){\partial\bar z_\Phi\over\partial y}
= {\chi y\over\xi^2}\bar z_\Phi
+ \left(\lambda + {4\over3}H\right)
\bar z_\Psi.
 \end{equation}

It is straightforward to integrate this equation using
an integrating factor, to obtain
 \begin{equation}
\bar z_\Phi(y) =
- {\lambda + {4\over3}H\over 2+\lambda y} \int_y^\infty
dy' \left(2+\lambda y'\over2+\lambda y\right)
^{n-1} e^{-\chi(y'-y)/\lambda\xi^2}
\bar z_\Psi(y'),
\label{zbarsol}
 \end{equation}
 where
 \begin{equation}
n = {2\chi\over\lambda^2\xi^2}.
\label{ndef}
 \end{equation}

As we noted earlier, the exponential
Ansatz breaks down near $y=0$, but if we ignore that
for the moment, we can estimate the value of
$\bar\alpha = - \bar z_\Phi(0)$, as
 \begin{equation}
\bar\alpha \approx {\lambda +
{4\over3}H\over 2} \int_0^\infty
dy \left(1+{\lambda y\over2}\right)
^{n-1} e^{-\chi y/\lambda\xi^2}
\bar z_\Psi(y).
\label{ab1}
 \end{equation}

For large values of $ y$, $\bar z_\Psi$ is expected to
fall off exponentially.  The model described in Section
2 suggests that $\bar z_\Psi\approx we^{-B y}$,  with
$B=w/\bar\xi$.  In that case, $\bar z_\Phi$ is expressible
in terms of the incomplete gamma function:
 \begin{equation}
\bar z_\Phi( y) =  - {w(\lambda+{4\over3}H)\over\lambda}
x^{-n} e^{x-By} \Gamma(n,x)
\label{zbarig}
 \end{equation}
where
 \begin{equation}
x = \left({B\over\lambda}+{n\over2}\right)
(2+\lambda y).
 \end{equation}
Expanding for large $x$, the leading term is
 \begin{equation}
\bar z_\Phi(y) \approx - {(\lambda
+ {4\over3}H)w e^{-B y}\over(B+{1\over2}n\lambda)
(2+\lambda y)}.
\label{zbarbigx}
  \end{equation}

As a consistency check, we may substitute this
solution into the equation (\ref{zbarPsid}) for
the large-$y$ $\Psi$ distribution and verify that
it can be satisfied with a reasonable form of
the unknown velocity-distortion function $v(y)$.
In fact, we find for the leading approximation
  \begin{equation}
v(y) = 3\left(w\lambda + {\chi\bar\xi\over\xi^2}
\right)y e^{-By},
  \end{equation}
which seems entirely reasonable.

For $y \gg 1/\lambda$, $-\bar z_\Phi$ is small
compared to $\bar z_\Psi$.  But as $y$ falls it
grows rapidly.  For moderate values of $y$ the two
are of comparable magnitude,
assuming that $1/\lambda$ is of the same order as
$\xi$ and $\bar\xi$.

We know from the simulations that even at $y=0$,
$\bar z_\Phi$ does not become large \cite{BB90,AS90}:
 \begin{equation}
\bar z_\Phi(0) = - \bar\alpha \approx -0.14
 \end{equation}
(in the radiation-dominated era).
Hence it is reasonable to assume for all values
of $y$ that $\bar z_\Phi \ll 1$.  This does not
necessarily mean that the exponential Ansatz is
valid (a point we shall return to shortly), but
so long as it is we can still use
(\ref{zbard}), with $(\overline{z^2})_\Phi
\approx {1\over3}$.  However, as $y\to 0$, we can no
longer use (\ref{zbarX}); instead, we have
both $\bar z_\Psi\to 1$ and $\bar z_{\rm X}\to 1$.
The net effect is that in the small-$y$ region,
the exponent $n - 1$ in (\ref{zbarsol}) becomes
$n$.  But since $\lambda y$ and $\lambda y' \ll 2$ in
that region, the effect is minimal.  The expressions
obtained above should still give a good approximation
to $\bar z_\Phi$.

\subsection{Behaviour near $y=0$}

To get at least a rough estimate of the
value of $\bar z_\Phi(0) = -\bar\alpha$,
let us first assume that $\lambda\xi \gg 1$ and
$\lambda\bar\xi \gg 1$.  Then in (\ref{zbarig})
both $x$ and $n$ are small.  The leading term
in the expansion of $\Gamma(n,x)$ for small $x$
and $n$ yields
 \begin{equation}
\bar\alpha \approx w\left(1 + {4H\over3\lambda}
\right)\ln(\lambda\bar\xi)\qquad(\lambda\bar\xi\gg1).
\label{albarsmx}
 \end{equation}
  This is, however, much
too large, clearly inconsistent with our assumption that
$\bar z_\Phi$ always remains small.

It is perhaps more plausible to assume that
$\lambda\xi \ll 1$ and $\lambda\bar\xi \ll 1$.
Then {\it both} $x$ and $n$ are large, and roughly
equal, and the
asymptotic form of the incomplete gamma function
gives
 \begin{equation}
\bar\alpha \approx {w\over2}\sqrt{\pi\over\chi}
\left(1 + {2H\over3\lambda}
\right)\lambda\xi\qquad(\lambda\xi\ll 1).
\label{albarbigx}
 \end{equation}

Although $\bar z_\Phi$ remains small, the exponential
Ansatz is not in fact a good approximation near $y=0$,
because the distribution $\Psi$ becomes so sharply
peaked near $z=1$.  The distribution functions in
the last two terms of (\ref{dPhitot}) are of course
both normalized, but the second one is negligibly
small over most of the angular range, becoming very
large near $z=1$.  Thus, while a linear approximation
to $\Phi$ remains good for most values of $z$, near
$z=1$ it becomes very poor.  As $y\to0$, $\Phi$
acquires a deep hole in the forward direction.
For this reason, our estimate of $\bar z_\Phi(0)$
requires some correction.

To estimate the likely size of the effect, let us
assume that
the exponential Ansatz is at least qualitatively
reasonable down to
values of $y$ of order $\zeta$.  So when we come to
consider the equation (\ref{dPhitot})
in the region of small $y$
we can use the exponential (or indeed linear)
form as our initial condition at $y\sim\zeta$.

When $y \ll\zeta$ we
can use the approximation (\ref{psismy}) for $\Psi$,
with $\bar z_\Psi \approx 1 - (y/\zeta)$.  Then our
equation (\ref{dPhitot}) for $\Phi$ becomes
 \begin{equation}
(2+\lambda  y){\partial\Phi\over\partial y}
= -2H\left(1 - {y\over\zeta}\right)
{\partial\over\partial z}\left[
(1-z^2) \Phi\right]
- {\chi y\over\xi^2}(1-\Phi)
- \lambda\Phi + \lambda N{2\zeta\over y}
e^{-(1-z)\zeta/y}\Phi.
 \end{equation}

It is convenient to change from $y$ to $k \approx
\zeta/y$ as the independent variable.  The equation
involves three small parameters, $\lambda\zeta$,
$H\zeta$ and $\chi\zeta^2/\xi^2$.
Since the last of these is very small indeed, the
intercommuting term will
give a very small contribution in this region.  Moreover,
we expect $H\zeta \ll \lambda\zeta \ll 1$, so the
contribution of the stretching term is also likely
to be small.  Neglecting these terms, and also
neglecting $\lambda y$ in comparison with 2, we find
\begin{equation}
2k^2{\partial\Phi\over\partial k}
=  \lambda\zeta(1 - 2Nk
e^{-k(1-z)})\Phi.
\label{dPhismy}
\end{equation}

We now seek to solve this equation, starting with some
initial value $k_0$ of $k$, at which we assume
that $\Phi$ has the exponential form (\ref{phidef}),
with a value $b_0$ of $b$.  Since $\lambda\zeta$ is small, we
may adopt a perturbation approach, taking as our
zero-order solution, $\Phi_0(k,z)$ = $\Phi(k_0, z)$,
independent of $k$.  Correspondingly, the zero-order
value of $N$ is given by
 \begin{equation}
N_0^{-1} = \int_{-1}^1 {dz\over 2} 2ke^{-k(1-z)}
\Phi(k_0, z) \approx \Phi(k_0,1) = {b_0 e^{-b_0}
\over \sinh b_0}.
 \end{equation}
Now we substitute this into the right hand side of
(\ref{dPhismy}) and integrate, obtaining in first order

 \begin{eqnarray}
\Phi(k,z) &=&
\Phi(k_0,z) \exp\left(
{\lambda\zeta\over 2k_0} - {\lambda\zeta\over 2k}
-\lambda\zeta N_0
\int_{k_0}^k {dk'\over k'} e^{-k'(1-z)}\right)
\nonumber\\
&=&
\Phi(k_0,z) \exp\left(
{\lambda\zeta\over 2k_0} - {\lambda\zeta\over 2k}
+ \lambda\zeta N_0
[{\rm Ei}\bigl(-k_0(1-z)\bigr)-
{\rm Ei}\bigl(-k(1-z)\bigr)]\right),
\label{Phismy}
 \end{eqnarray}
 where ${\rm Ei}$ is the exponential integral.

It is interesting to examine the special case $z=1$.
At that point, we have
 \begin{equation}
\Phi(k,1) \approx N_0^{-1} \exp\left(
{\lambda\zeta\over 2k_0} - {\lambda\zeta\over 2k}
\right) \left(k\over k_0\right)^{-\lambda\zeta N_0}.
 \end{equation}
Note that $\Phi(k,1)$ approaches zero as $k\to\infty$
(or $y\to0$), but very slowly.

The most interesting limit of course is $k\to\infty$.
Strictly speaking, in this limit the first-order
approximation in $\lambda\zeta$ breaks down, but in fact it
does so only at such large values of $k$ that $\partial
\Phi/\partial k$ is already negligibly small.  There seems
no need to go beyond first order.  For $z\ne1$, we have
in that limit
 \begin{equation}
\Phi(k=\infty, z) = \Phi(k_0, z)
 \exp\left(
{\lambda\zeta\over 2k_0}
+ \lambda\zeta N_0
{\rm Ei}\bigl(-k_0(1-z)\bigr)\right).
\label{Phiy0}
 \end{equation}
 The behaviour of this function near $z=1$
is given by
 \begin{equation}
\Phi(k=\infty,z) \approx N_0^{-1} e^{-b_0(1-z)}
\exp\left(
{\lambda\zeta\over 2k_0}\right) [k_0(1-z)]^{\lambda\zeta N_0},
\qquad (1-z \ll 1).
 \end{equation}
This clearly exhibits the expected sharp dip near
$z=1$: the last factor vanishes at that point, but is
close to unity over most of its range.

We are particularly interested in the average value
$\bar z_\Phi(y=0) = -\bar \alpha$, obtained by
multiplying (\ref{Phiy0}) by $z$ and integrating.
Since $\Phi(y=0,z)$ is a product of two factors, each
of which separately gives a small value of $\bar z$,
the two effects are approximately additive, and we may
write
 \begin{equation}
\bar \alpha\approx \bar\alpha_{\rm lin} + \bar\alpha_{\rm nl},
\label{absum}
 \end{equation}
where the subscripts stand for `linear' and
`non-linear'.    Here the linear contribution
$\bar\alpha_{\rm lin}$ is the value given by (\ref{albarsmx}),
(\ref{albarbigx}), or something in between,
while the non-linear term
$\bar\alpha_{\rm nl}$ comes from the second
factor in (\ref{Phiy0}).  To first order in $\lambda\zeta$,
it is
 \begin{equation}
\bar \alpha_{\rm nl} = - (\bar z_\Phi)_{\rm nl}
\approx {\lambda\zeta\over2}.
\label{abnldef}
 \end{equation}

\section{ANGULAR CORRELATION AND LOOP FORMATION}

We are now in a position to return to the calculation
of the probability of loop production,
$\Theta({\bf r},l)$. The provisional formula
(\ref{Th2}) is wrong because it assumes that
the probability distributions of the left- and
right-moving segments are independent.  It should
be replaced by (\ref{Th4}).

To complete our programme of expressing all the terms in
the equation for $\partial p[{\bf r}(l)]/\partial t$
in terms of $p$ itself, we have to reexpress
the joint probability here
in terms of individual probabilities.

\subsection{Small loops}

Let us first consider the case of
very small loops. We shall deal separately with larger
loops in the  next subsection.

In the case of small loops, the essential effect is
due to the angular correlation between ${\bf p}$ and ${\bf q}$
vectors. To be completely correct, we should consider the
joint probability distribution of all the
${\bf p}$ and ${\bf q}$ vectors forming this
section of string.  This would obviously be a
very complicated object; we are not in a position to
deal with it.  However, in the case of small
loops, the internal correlation between ${\bf p}$
vectors at different points is very strong (as is that
between ${\bf q}$ vectors), so
it seems reasonable, in order to represent the effect
of the ${\bf p}$-${\bf q}$ correlation,
to choose a single representative
vector from each class.  We choose the pair for which
the effect is strongest, namely the vectors at the
mid-points of the segments, namely ${\bf p}(l/2)$ and
${\bf q}(l/2)$.  (We have considered as an alternative
averaging over the chosen position; this makes little
difference.)

Thus we take
 \begin{equation}
p[{\bf r}(l);{\bf r}(l)] \approx
\int {d^2{\bf p}\over4\pi}\,
{d^2{\bf q}\over4\pi}\,
\Phi(0,{\bf p}{\cdot}{\bf q})\,
p[{\bf r}(l)|{\bf p}(l/2)]
\,p[{\bf r}(l)|{\bf q}(l/2)],
\label{prrsl}
 \end{equation}
where $p[{\bf r}(l)|{\bf p}(l/2)]$ is the
probability distribution of extension conditional
on the direction of the vector ${\bf p}$
at the  mid-point, $d^2{\bf p}$ denotes an
integration over the unit sphere and of course
$\Phi$ is the angular distribution function for
$z={\bf p}{\cdot}{\bf q}$ evaluated at
the point where the two vectors meet, namely
$y=0$.  It is reasonable to assume that, apart
from their mutual correlation, the ${\bf p}$
and ${\bf q}$ vectors are uniformly distributed
on the sphere.

The Gaussian approximation should be valid, both
for the conditional probabilities and for the
loop-production function $\Theta({\bf r},l)$,
since there is an effective cutoff in the region of very
small loops due to the behaviour of the  angular
distribution function and the volume factor.  Then
$p[{\bf r}(l)|{\bf p}(l/2)]$ is the
Gaussian distribution with
appropriate values of $\bar{\bf r}$ and
$\overline{{\bf r}^2}$.
By (\ref{rpexp}),
 \begin{equation}
\bar{\bf r} = S{\bf p},
 \end{equation}
where
 \begin{equation}
S = \overline{{\bf r}{\cdot}{\bf p}} = K'(l/2).
 \end{equation}
The value of $\overline{{\bf r}^2}$ is unaffected
by ${\bf p}$: $\overline{{\bf r}^2} = K(l)$.  This means
of course that the variance of ${\bf r}$ is reduced:
 \begin{equation}
\hat K(l) \equiv \overline{{\bf r}^2}
- \bar{\bf r}^2 = K - S^2.
 \end{equation}

It is worth remarking that for moderately small
values of $l$ the Gaussian Ansatz is a much better
approximation for the conditional probability
$p[{\bf r}(l)|{\bf p}(l/2)]$ than it is for the
unconditional probability $p[{\bf r}(l)]$.  For
small $l$, the probability distribution is of course
concentrated near the sphere
${\bf r}^2 = l^2$ and is nothing like a Gaussian
centred at ${\bf r}={\bf 0}$.  However,
the Gaussian approximation to $p[{\bf r}(l)|{\bf p}(l/2)]$ is
centred near that sphere and has much smaller variance,
since for small $l$
 \begin{equation}
S\approx l - {l^2\over4\zeta},\qquad
\hat K \approx {l^3\over6\zeta}\qquad
({\rm small}\;l).
 \end{equation}

We can now compute $\Theta$, or the variance function
$Q$ in the Gaussian approximation.

Substituting (\ref{prrsl}) into (\ref{Th4}),
we obtain
 \begin{eqnarray}
\Theta({\bf r},l)p[{\bf r}(l)] &\approx&
\Lambda(l)\left(3[K(l)+Q(l)]\over2\pi
K(l)Q(l)\right)^{3/2}
e^{-3{\bf r}^2[1/2K(l)+1/2Q(l)]}
\nonumber\\
&\approx&
2e^{-\chi l t/\xi^2}\Delta({\bf r},l)
\int {d^2{\bf p}\over4\pi}\,
{d^2{\bf q}\over4\pi}\,
\Phi(0,{\bf p}{\cdot}{\bf q})\,
p[{\bf r}(l)|{\bf p}(l/2)]
\,p[{\bf r}(l)|{\bf q}(l/2)].
\nonumber\\
&&
\label{Th5}
  \end{eqnarray}
It is easy to carry out the integrations over
${\bf p}$ and ${\bf q}$, leaving only a single integration which
may be written
  \begin{eqnarray}
\Theta({\bf r},l)p[{\bf r}(l)] &\approx&
2e^{-\chi l t/\xi^2}\Delta({\bf r},l)
\left(3\over2\pi\hat K\right)^3
e^{-3({\bf r}^2+S^2)/\hat K}
\nonumber\\
&&\times{\hat K\over3Sr}
\int_0^1 du\,\Phi(0,2u^2-1)
\sinh\left({6Sr\over\hat K}u\right).
 \end{eqnarray}
  Without having a specific form for
$\Phi$, it is not possible to proceed further.
However, we can get a good idea of the likely effect by
assuming that $\Phi$ has a sharp step-function cutoff,
{\it i.e.}, $\Phi = 0$ for $z > z_0 \equiv
1-2\bar\alpha_{\rm nl}$ and $\Phi =$ constant for $z<z_0$.
We then find
 \begin{equation}
\Theta({\bf r},l)p[{\bf r}(l)] \approx
2e^{-\chi l t/\xi^2}\Delta({\bf r},l)
\left(3\over2\pi\hat K\right)^3
e^{-3({\bf r}^2+S^2)/\hat K}
{\sinh^2 X\over X^2},
 \end{equation}
with
 \begin{equation}
X=\sqrt{1-\alpha_{\rm nl}}{3Sr\over\hat K}.
 \end{equation}

Using (\ref{delta0def}), we can now perform the
integration over ${\bf r}$ in (\ref{Lamdef}) to find $\Lambda$.
For small $l$, $S^2/\hat K \approx 6\zeta/l$,
so there is a strong exponential cutoff for
$l <\zeta$.  Thus we can legitimately set
$\Delta({\bf r},l) \approx \Delta_0$.  Then we obtain
 \begin{equation}
\Lambda \approx {\Delta_0\over
4\pi(1-\bar\alpha_{\rm nl})}
\left(3\over\pi \hat K\right)^{1/2}
{1\over S^2} \left(e^{-3\bar\alpha_{\rm nl} S^2/\hat K}
- e^{-3S^2/\hat K}\right).
\label{Lam3}
 \end{equation}

\subsection{The intermediate-scale region}

The formula (\ref{prrsl}) incorporates our model of
the ${\bf p}$-${\bf q}$ correlation which, as we have seen,
provides a very effective cutoff for values of $l$
of order $\zeta$ or less.  However, it would still
predict an impossibly large value of $\lambda$.  If we
substitute from (\ref{Lam3}) into (\ref{lamdef}), this
arises from the contribution of the intermediate region,
where
 \begin{equation}
\zeta \ll l \ll\bar\xi.
\label{intl}
 \end{equation}
 In this region, $\hat K \sim w(1-w)l^2$,
while $S^2/\hat K \sim w/(1-w)$
is of order unity, so the integrand in (\ref{lamdef})
behaves like $1/l^2$.  Thus we find  $\lambda \sim 1/\zeta$,
up to a constant of order one.   This would mean an
extremely rapid decrease of $L$, on a time-scale of
order $\zeta$, which is clearly not  consistent with the
results of the simulations.

The explanation for this discrepancy again lies in the
angular correlation effect, but of longer segments of
string, not merely individual ${\bf p}$ and ${\bf q}$
vectors.

Think of a section of string of length $l$ and
extension ${\bf r}$, and suppose that $\zeta \ll l \ll \bar\xi$.
In other words, the section contains many kinks, but
viewed on a large scale it is likely to be fairly
straight, {\it i.e.}, $|{\bf r}|$ is a sizable fraction of $l$.
Now consider the collection of ${\bf p}$ vectors on this
section.  Their ensemble average is of course $\langle
{\bf p}\rangle = {\bf r}/l$.  So the distribution of ${\bf p}$
vectors will be strongly skewed, concentrated into a
cone around the direction of ${\bf r}$.

The ${\bf q}$ vectors on the corresponding right-moving
section will also be concentrated in a cone, around the
direction of ${\bf r}_{\rm r}$.  In most cases, the two
cones will not overlap much, and rather few loops will
form.  On the other hand, where the cones do overlap,
many loops will form.  Such a section will disappear
rapidly in a burst of loop formation.  After a short
time, regions where the distributions overlap will be
rare.

Of course, regions of overlap are continually being
reformed, as sections of string meet new partners.  In
particular, the intercommuting process brings together
segments of string that had been far apart and are
therefore more or less uncorrelated.  In some fraction
of cases, depending on the size of the cones, these
segments will have overlapping distributions, so a
fresh burst of loop formation will be triggered.  This
burst phenomenon has indeed been observed in the
simulations.

It is important to note that the angular concentration
effect will be even more pronounced for somewhat larger
loops.  One may think of the string section as composed
of short straight segments of length $\sim\zeta$, with
randomly varying orientation within the cone around
${\bf r}$.  Thus their transverse extensions are
essentially a two-dimensional random walk.  If we
select a length $n\zeta$, its overall
extension  will be close to $n\zeta{\bf r}/l$, with a
transverse spread proportional to $\sqrt n \zeta$.  Hence
the angular distribution of such sections will be
concentrated in a cone whose angle is reduced by a
factor $1/\sqrt n$ compared to the cone of ${\bf p}$ vectors.

The question we want to answer is: how does the loop
formation probability depend on the three scale
lengths $\zeta, \bar\xi, \xi$?

In approaching this question, let us begin by
considering an initial state in which all three
scales are comparable in magnitude, as would be
expected shortly after the string-forming phase
transition: $\zeta\sim\bar\xi\sim\xi < t$.  In this case, the
upper cutoff of the loop formation integral, at
$\xi^2/\chi t$, may well be smaller than the lower
cutoff, $\zeta$.  What this means is simply that almost
all loops formed reconnect to the network.  There is
then no reason why either $\zeta$ or $\bar\xi$ should grow
rapidly.  If we consider only the stretching terms, we
expect  (in the radiation era)

 \begin{equation}
{\dot\xi\over\xi} = {3- \alpha\over2}
H = {3 - \alpha\over4t}.
 \end{equation}
Here $\alpha$ may be dependent on the ratios of length
scales, but its value is presumably small, say around
0.2.  This implies that $\xi\propto t^{0.7}$.  Thus
we reach a regime where $\zeta\sim\bar\xi\ll\xi < t$.  The
upper cutoff does grow, if only rather slowly:
$\xi^2/\chi t \propto t^{0.4}$.

The next stage is easy to describe, at least
qualitatively.  Large numbers of small loops start to
form, without reconnection, increasing the rate of
growth of $\xi$.  But also $\bar\xi$ starts to grow.  This
is because of the selective nature of loop formation.
Some sections of string will be relatively straight.
On those, the distributions of ${\bf p}$ and ${\bf q}$ vectors
will be confined to cones.  In a few such cases, the
cones may overlap, leading to rapid disappearance; in
most, they will not.  By contrast, on the more curled
sections of string, ${\bf p}$ or ${\bf q}$ vectors will
readily find partners.  Those sections will disappear
too, though not perhaps quite as rapidly as where
there are overlapping cones.  The net result will be to
eliminate selectively the more curled sections of
string, leading to growth of $\bar\xi$.

One consequence of this that will be important
later is that the function $\Theta({\bf r},l)$
representing the probability of loop formation will be
more concentrated towards ${\bf r}={\bf 0}$
than would otherwise have been expected.  In other
words, the variance function $Q$ defined in
(\ref{Qdef}) will be smaller than the original forms
(\ref{Th1}) and (\ref{delta0def}) would suggest.

Although $\bar\xi$ starts to grow, there is still no reason
for $\zeta$ to do so.
Every intercommuting event and every
loop formation introduces new kinks, keeping down the
average interkink distance.  (As we shall see later,
there is good reason to believe that $\zeta$ does
eventually start to grow, but only when gravitational
radiation becomes important.)

Thus, within a few expansion times, we may expect to
reach a regime where $\zeta\ll\bar\xi\ll\xi < t$.  How, in
such a regime, does the loop formation rate depend on
the various length scales?  In other words, what
correction factor do we need to apply to the rate
(\ref{Th4}) calculated neglecting the angular
correlation effects?

Consider a particular loop size $\l$ within the range
$\zeta\ll l \ll \xi^2/\chi t$.  We also assume that
$l\ll \bar\xi$.  Typically then a segment of length $l$
will be part of a longer section that is roughly
straight on a scale of order $\bar\xi$, with many kinks
separated by distances of order $\zeta$.  The segments of
length $l$ will be concentrated around this overall
direction.  Unless this is a region where there is a
burst of loop formation, the corresponding
right-moving segments will be concentrated in a
similar, but essentially non-overlapping, cone ---
simply because regions where cones overlap quickly
disappear.

In these circumstances, loop formation will occur only
when segments meet new partners with a different
overall direction.  This happens for  two reasons ---
intercommuting and the steady progression of left- and
right-movers.

Consider first intercommuting.  The probability that a
segment of length $l$ experiences intercommuting
within a time interval $\delta t$ is $\chi l \delta
t/\xi^2$, or, to put it another way, the average time
between such events is $\xi^2/\chi l$.

The steady progression will bring a given
segment alongside ones with a different overall
orientation when the left- and right-movers have moved
relatively by a distance of order $\bar\xi$, {\it i.e.}, after a
time $\bar\xi/2$.  In the regime we are presently
considering, this is a much shorter time, so the steady
progression is the important effect.  (The
intercommuting time would be shorter only for segments
well above the upper cutoff length.)  It is therefore
reasonable to ignore the intercommuting effect.

It is important to realise that this does not mean
intercommuting plays no role; quite the contrary.  What
we are saying is that the {\it direct\/} effect of
intercommuting {\it within\/} the chosen segment is
unimportant.  The {\it indirect\/} effect of nearby
intercommuting is one of the things that keeps $\bar\xi$
down, and is in fact clearly crucial.

If the orientations of the segments of length $l$ were
random, a given segment would meet another of
essentially different orientation every time it moved
on by a distance $l$, {\it i.e.}, after a time $l/2$.  As it
is, because of the long-range correlation, it will meet
a segment with an essentially different orientation
only after a time $\bar\xi/2$.    On this
ground, the rate of loop formation should be suppressed
by a factor of approximately $l/\bar\xi$ (for $l<\bar\xi$; there
is no suppression for $l>\bar\xi$).

This is not the end of the story, however.  We have to
consider what happens when a new segment is
encountered.  The new right-moving section of string
may have any orientation relative to the left-moving
section.  We may assume that the relative orientation
is random.  Thus if $\alpha$ is the angle between the two,
then $\cos\alpha$ should be uniformly distributed between
$-1$ and $1$.

As we saw, the left-moving segments of length $l$ lie
within a cone whose semi-vertical angle $\beta$ is
proportional to $1/\sqrt n$, where $n$ is, roughly
speaking, the number of sizable kinks on the segment,
namely $n\approx l/\zeta$.  The solid angle within the cone
is of order $\pi\beta^2\sim\pi\zeta/l$.

Now clearly, if $\alpha$ is significantly larger than
$\beta$, then our chosen segment is very unlikely to
meet a matching partner.  On the other hand, in the
relatively rare cases in which $\alpha<\beta$, loop
formation is extremely probable.  In other words, we
may expect {\it another\/} suppression factor, roughly
equal to the solid angle of the cone divided by $4\pi$,
namely $\zeta/4l$.

Putting these two suppression factors together, we
obtain an additional factor in $\Lambda(l)$ of about
$\zeta/4\bar\xi$.  We now insert the expression for $\Lambda(l)$,
given by (\ref{Lam3}) with this factor, into the formula
(\ref{lamdef}) for the overall loop-formation rate
$\lambda$.  Here we have to integrate from the lower limit
$\zeta$ to the upper limit $\xi^2/(\chi t)$.  Since
$\Lambda(l) \propto 1/l^3$, the integral behaves like $\int
dl/l^2$, and it is the lower limit that dominates ---
as must be true if most loops produced are indeed
small.  Hence the integral is of order $1/\zeta$.  Taking
account of the extra suppression factor, we see from
(\ref{Lam3}) and (\ref{lamdef}) that
  \begin{equation}
\lambda = {E\over\bar\xi},
  \end{equation}
where $E$ is a dimensionless parameter rather less than
unity.

Note that the index $n$ defined in (\ref{ndef}) is
  \begin{equation}
n = {2\chi\over E^2} {\bar\xi^2\over\xi^2}.
  \end{equation}
So when $\bar\xi\ll\xi$, $n$ may be small, but for
$\bar\xi\sim\xi$, it will be of order unity or even
somewhat larger.

\subsection{Rates of change of $\xi$ and $\bar\xi$}

We now take up the calculation of the effect of
loop formation on the rates of change of the length
scales in the problem.

First, we consider $\xi$.  It is of course defined in
terms of $L$ by (\ref{xidef}), from which it follows that
  \begin{equation}
{\dot\xi\over\xi} = {3H\over2} - {\dot L\over2L}.
\label{xid}
  \end{equation}
Thus the rate of
change of $\xi$ due to loop formation is
 \begin{equation}
{\dot\xi_{\rm loops}\over\xi} = {\lambda\over2}
= {E\over2\bar\xi},
\label{xidllf}
 \end{equation}
where the dimensionless function $E$ is given by
 \begin{equation}
E\left({\zeta\over\xi},{\bar\xi\over\xi},
{t\over\xi}\right) = {\lambda\bar\xi}
= \bar\xi \int_0^\infty dl\,l\Lambda(l).
\label{Edef}
 \end{equation}

For the other two scales, we have to examine the rate
of change of the variance function $K$.  The Gaussian
approximation should be valid, since there is an
effective cutoff in the region of very small loops due
to the behaviour of the  angular distribution function
and the volume factor.

It is then straightforward to derive an expression for
the rate of change of $K$ by performing the integrations
over ${\bf r}_1$ and ${\bf r}$ in (\ref{pdll}) explicitly.

In the Gaussian approximation, the joint probabilities
are given by expressions of the form (\ref{p12})
with ${\bf K}$  replaced by the covariance matrix
(\ref{K01}) and $\overline{{\bf r}{\cdot}{\bf r}_1}$
given by (\ref{r10exp}).

For the first term of (\ref{pdll}), the appropriate
covariance matrix is
 \begin{equation}
{\bf K} = \biggl[
\begin{array}{cc}
K_1&K_a\\K_a&K_b
\end{array}
\biggr],
\label{Kpos}
 \end{equation}
where
 \begin{eqnarray}
K_1 &=& K(l_1),
\nonumber\\
K_a &=& \case{1}/{2}[K(l_1+y_0) + K(l+l_1-y_0)
- K(y_0) - K(l-y_0)],
\nonumber\\
K_b &=& K(l+l_1),
 \end{eqnarray}
while for the second it is
 \begin{equation}
{\bf K} = \biggl[
\begin{array}{cc}
K_1&K_c\\K_c&K
\end{array}
\biggr],
 \end{equation}
where
 \begin{eqnarray}
K_c &=& \case{1}/{2}[K(l_1+y_0) + K(l-y_0)
- K(y_0) - K(l-l_1-y_0)],\nonumber\\
K &=& K(l).
 \end{eqnarray}

The exponent in the first term of the integrand in
(\ref{pdll}) contains the inverse of (\ref{Kpos})
plus a single-entry contribution from the extra
factor $\Theta({\bf r}_1,l_1)$.  In other words, the
inverse of ${\bf K}$ is replaced by the inverse of
a new matrix {\bf L}, given by
 \begin{equation}
{\bf L}^{-1} = {\bf K}^{-1} +
{1\over Q_1}\biggl[
\begin{array}{cc}
1&0\\0&0
\end{array}
\biggr],
 \end{equation}
with $Q_1 = Q(l_1)$. Thus
 \begin{equation}
{\bf K}\,{\bf L}^{-1} = {1\over Q_1}\biggl[
\begin{array}{cc}
Q_1+K_1&0\\K_a&Q_1
\end{array}\biggr],
\end{equation}
whence we find
\begin{equation}
{\bf L} = {1\over Q_1+K_1} \biggl[
\begin{array}{cc}
Q_1K_1\quad&Q_1K_a\\
Q_1K_a\quad&K_b(Q_1+K_1)-K_a^2
\end{array}
\biggr].
 \end{equation}

It is now straightforward to take moments of
(\ref{pdll}) to find $(\partial K/\partial t)_{\rm loops}$.  When
we perform the ${\bf r}$ and ${\bf r}_1$ integrations,
we obtain a determinantal factor, which essentially
cancels the normalization constants, and a factor
corresponding to the expectation value of ${\bf r}^2$ in the
appropriate distribution.

Let us consider the expectation value factor.  In
performing the ${\bf r}$  and ${\bf r}_1$ integrations
in the first term, it is simplest to change variables
from ${\bf r}$  to ${\bf r}' = {\bf r} + {\bf r}_1$, so that
${\bf r}^2$ becomes $({\bf r}'-{\bf r}_1)^2$.  Hence the
appropriate expectation value comprises a sum
of elements of the matrix {\bf L}.  In the
second term, we have only a single element contributing.

In this way we obtain
 \begin{eqnarray}
\left(\partial K\over\partial t\right)_{\rm loops}
&=& \int_0^\infty dl_1\,
\Lambda(l_1)\biggl[l\left(
K_b-K + {Q_1K_1\over Q_1+K_1}
\right)
\nonumber\\
&&-{2Q_1\over Q_1+K_1}
\int_0^l dy_0\,K_a
-{1\over Q_1+K_1}
\int_0^l dy_0\,K_a^2
+{1\over Q_1+K_1}
\int_{-l_1}^l dy_0\,K_c^2
\biggr].
\nonumber\\
&&
\label{Kdllf}
 \end{eqnarray}

 From (\ref{xbd}) and (\ref{Kdllf}),
we now obtain
 \begin{equation}
{\dot{\bar\xi}_{\rm loops}\over\bar\xi}
= {1\over2\bar\xi}I\left({\zeta\over\xi},{\bar\xi\over\xi},
{t\over\xi}\right),
\label{xbdllf}
 \end{equation}
where $I$ is another dimensionless function of the
scale ratios defined by
 \begin{equation}
I = \int_0^\infty dl_1\,
\Lambda(l_1)\left(2\bar\xi l_1 +
{Q_1(K_1 - 4 \bar\xi l_1)\over
Q_1+K_1}\right).
\label{Idef}
 \end{equation}

\subsection{Rate of change of $\zeta$}

Estimating the rate of change of $\zeta$ is less
straightforward.   The region of the integrand
where $l_1$ is very small is
strongly suppressed by the $\Lambda(l_1)$
factor, so we need not be particularly concerned
with the behaviour of the remaining factors in this
region.  However, the same is not true of
the region where the length $l$ of our
original segment becomes small.
In the small-$l$ case there is a special feature
that requires separate attention.

We noted in Section \ref{sec-smalll} that for small values
of $l_1$ the conditional expectation value of ${\bf r}_1^2$
is no longer given by the expression (\ref{r1sqexp})
but rather by (\ref{r1sqsmall}); it reduces to the
{\it un\/}conditional expectation.  The same
argument applies to the conditional expectation value of
${\bf r}^2$ for small $l$.  It follows that in the
contribution to (\ref{Kdllf}) arising from the  second
(negative) term of (\ref{Lpdll}), the integrand should
in that limit be simply  $K \approx l^2$,
rather than  $K - K_c^2/(Q_1+ K_1)$.

 The same argument applies to the first (positive)
term.  It is true that neither $l_1$ nor $l+l_1$ goes
to zero in the small-$l$ limit, so the mean values of
${\bf r}_1^2$, ${\bf r}'{}^2$ or ${\bf r}_1\cdot{\bf r}'$ individually
should be well approximated by the Gaussian form.
However, we are interested in the expectation value of
${\bf r}^2$, which is a small difference of these large
quantities.  Clearly, it too must approach $l^2$ in the
limit $l\to 0$.

If we simply set the mean value of ${\bf r}^2$
everywhere equal to $K(l)$, the entire contribution
would cancel out.  However, this may not be quite
correct.

Suppose a loop is formed and that we choose a very
small segment of length $l$ on the loop.  What is the
variance of its extension?  Clearly the leading term
for small $l$ is $l^2$, but we could easily have
 \begin{equation}
\overline{{\bf r}^2_{\rm loop}} =
l^2 - {(1+k)l^3\over3\zeta},
 \end{equation}
with a value of $k\ne0$.  Indeed, it seems likely that
loops are generally kinkier on small scales than long
strings, in which case we would expect $k>0$.

In this case, we easily find
 \begin{equation}
{\dot\zeta_{\rm loops}\over\zeta}
= k\lambda.
\label{zedllf}
 \end{equation}

\section{STRETCHING}

 To complete our central task of deriving the
various terms in the evolution equations for $p[{\bf r}(l)]$
and $K(l)$, we have to estimate the various parameters
and unknown functions that appear in them.
For the stretching terms, we need to examine the
parameters   $\alpha$ and $\beta$, which are defined by
(\ref{aldef1}) and (\ref{betadef1}).

To match the notation of the preceding section, we
denote the integration variable in these expressions
by $y_0$.  Note that the expectation values
appearing here are those conditional on
the extension ${\bf r}$ of our  chosen segment of length
$l$.

\subsection{Evaluation of $\beta$}

First consider $\beta$, given by (\ref{betadef1}).
To estimate the expectation value
$\langle{\bf q}(y_0)\rangle$ conditional on the
values of ${\bf r}$, we again use the equation of motion
for ${\bf q}$, (\ref{eqmotq}).  We could now derive an
equation, similar to that for $\Phi$ in the preceding
section,  but now for the complete angular
distribution, $\Phi[{\bf q}(y_1)]$ say, of ${\bf q}(y_1)$.

However, we can use a simplifying Ansatz.  The effect of
${\bf r}$ is important mainly  in the region of relatively
large values of $y_0-y_1$, in which it is reasonable to
use a generalization of the exponential Ansatz
(\ref{phidef}), namely
 \begin{equation}
\Phi[{\bf q}(y_1)] = {|{\bf b}|\over
\sinh|{\bf b}|} e^{{\bf b}{\cdot}{\bf q}},
 \end{equation}
where, as in (\ref{psibigy}),
 \begin{equation}
{\bf b} \approx 3\langle{\bf q}(y_1)\rangle.
 \end{equation}
Thus all we need is an equation for the expectation
value $\langle{\bf q}(y_1)\rangle$.

The required equation is a
simple generalization of (\ref{zbard}),
namely
 \begin{eqnarray}
(2+\lambda y){\partial\langle{\bf q}(y_1)
\rangle\over\partial y}
&=& H\langle{\bf p}(y_1) -
{\bf q}\,{\bf q}{\cdot}{\bf p}(y_1)\rangle
\nonumber\\
&&+{\chi y\over\xi^2}
\langle{\bf q}(y_1)\rangle
- \lambda\langle{\bf q}(y_1)\rangle
+ \lambda\langle{\bf q}(y_1)\rangle_{\rm X},
\label{qavd}
  \end{eqnarray}
with $y = y_0-y_1$ as before.  Here
$\langle{\bf q}(y_1)\rangle_{\rm X}$ denotes the
expectation value in the distribution ${\rm X}$.   Note
that because ${\bf p}(y_0)$ does not appear here, there
is only a single stretching term on the right hand side;
the factor of 2 that appeared in (\ref{zbard}) is absent.

Consider first the region of large $y$, where
the expectation values are small.  Within this region
we may write, as in (\ref{zbarX}),
  \begin{equation}
\langle{\bf q}(y_1)\rangle_{\rm X}
=\langle{\bf p}(y_1)\rangle
+\langle{\bf q}(y_1)\rangle
  \end{equation}
and replace $\langle{\bf p}(y_1) -
{\bf q}\,{\bf q}{\cdot}{\bf p}(y_1)\rangle$ by
${2\over3}\langle{\bf p}(y_1)\rangle$.
Then, as in (\ref{zbarsol}), we
find
  \begin{equation}
\langle{\bf q}(y_0)\rangle
\approx - {\lambda + {2\over3}H\over 2}
\int_0^{\infty} dy \left(1+{\lambda
y\over2}\right) ^{n-1} e^{-\chi
y/\lambda\xi^2} \langle{\bf p}(y_1)\rangle,
\label{qavsol}
  \end{equation}
where $n$ is again given by (\ref{ndef}).  The factor
$4\over3$ in (\ref{zbarsol}) is here replaced by
$2\over3$ because there is only a single stretching term.

The conditional expectation value
$\langle{\bf p}(y_1)\rangle$
which appears here was evaluated in the Gaussian
approximation in Section III.  It is given by
(\ref{pexp}).

As before, there will also be a non-linear
contribution to $\langle{\bf q}(y_0)\rangle$ arising from
the sharp hole in the angular distribution for
$z$ close to 1.  So long as the
contributions to this quantity are both reasonably small
(which appears to be the case), it should again be well
represented by a sum
 \begin{equation}
\langle{\bf q}(y_0)\rangle \approx
\langle{\bf q}(y_0)\rangle_{\rm lin} +
\langle{\bf q}(y_0)\rangle_{\rm nl}.
\label{qav}
 \end{equation}
Moreover, the non-linear contribution to
${\bf q}(y_0){\cdot}{\bf p}(y_0)$ is not much
affected by the value of ${\bf r}$, so it is reasonable to
set
 \begin{equation}
\langle{\bf q}(y_0)\rangle_{\rm nl} \approx
-\bar\alpha_{\rm nl}\langle{\bf p}(y_0)\rangle.
\label{qavnl}
 \end{equation}

Substituting from (\ref{qav}), (\ref{qavsol}) and
(\ref{qavnl}), we thus obtain
  \begin{equation}
\langle{\bf q}(y_0)\rangle =
- \bar\alpha_{\rm nl}\langle{\bf p}(y_0)\rangle
- {\lambda + {2\over3}H\over 2} \int_0^\infty
dy \left(1+{\lambda y\over2}\right)
^{n-1} e^{-\chi y/\lambda\xi^2}
\langle{\bf p}(y_0-y)\rangle.
\label{qexpG}
  \end{equation}
 Now, by (\ref{pexp}),
  \begin{equation}
\langle{\bf p}(y_1)\rangle =
{K'(y_1)+K'(l-y_1)\over2K(l)}
{\bf r}.
  \end{equation}
Thus,
(\ref{betadef1}) yields
  \begin{equation}
\beta(l) = \bar\alpha_{\rm nl} + {\lambda
+ {2\over3}H\over 4K(l)}
\int_0^l dy_0
\int_0^\infty dy\,[K'(y_0-y)+K'(l-y_0+y)]
\left(1+{\lambda y\over2}\right)^{n-1}
e^{-\chi y/\lambda\xi^2}.
  \end{equation}
 We may now perform the integration over $y_0$,
obtaining finally
  \begin{equation}
\beta(l) = \bar\alpha_{\rm nl} + {\lambda
+ {2\over3}H\over 4K(l)}
\int_0^\infty dy\,[K(l-y)+K(l+y)-2K(y)]
\left(1+{\lambda y\over2}\right)^{n-1}
e^{-\chi y/\lambda\xi^2}.
\label{betaG}
  \end{equation}
 (Recall that for negative values of the argument,
$K(y)$ is to be interpreted as $K(|y|)$.)
We note that at least in this
approximation $\beta$ is actually
a function of $l$ only, independent of {\bf r}.

Although we shall not require an explicit form, it is
interesting to note that if $K$ is assumed to have
the two-scale exponential
form (\ref{K2sc}), the integral (\ref{betaG}) can
be evaluated in terms of  the incomplete gamma
function.   (Note that to do this, one must split
the range of integration into separate ranges $0<y<l$
and $l<y<\infty$.)

\subsection{Evaluation of $\alpha$}

Now let us turn to the
evaluation of $\alpha$.  The analogue of (\ref{qavsol}) is
of course
 \begin{equation}
\langle{\bf p}{\cdot}{\bf q}(y_0)\rangle
\approx - {\lambda + {4\over3}H\over 2}
\int_0^\infty
dy \left(1+{\lambda y\over2}\right)
^{n-1} e^{-\chi y/\lambda\xi^2}
\langle{\bf p}(y_0){\cdot}{\bf p}(y_0-y)\rangle.
\label{pqavsol}
  \end{equation}
The factor of $2\over3$ is replaced here by $4\over3$
because, as in (\ref{zbarsol}), we now have to include the
effect of the rate of change of ${\bf p}(y_0)$.

Substituting from
(\ref{ppexp}) and adding a nonlinear contribution, we
then get as before
  \begin{eqnarray}
\langle{\bf p}{\cdot}{\bf q}(y_0)\rangle &=&
- \bar\alpha_{\rm nl} -
{\lambda + {4\over3}H\over 2}
\int_0^{\infty} dy\biggl\{
{1\over2}K''(y)
+ {1\over4} [K'(y_0)+K'(l-y_0)]
\nonumber\\
&&\times [K'(y_0-y)+K'(l-y_0+y)]
{{\bf r}^2 - K(l)\over K(l)^2}
\biggr\}
\left(1+{\lambda y\over2}\right)^{n-1}
e^{-\chi y/\lambda \xi^2}.
\label{alG}
  \end{eqnarray}

First, let us examine the mean value $\bar\alpha$, given
according to (\ref{aldef1}) by
  \begin{equation}
\bar\alpha l = - \int_0^l dy_0\,
\overline{{\bf p}{\cdot}{\bf q}}(y_0).
  \end{equation}
 From (\ref{alG}), it is obvious that
$\overline{{\bf p}{\cdot}{\bf q}}(y_0)$ (to which
the second term in the braces does not contribute) is in
fact  independent of $y_0$.  Hence we find
  \begin{eqnarray}
\bar\alpha = - \overline{{\bf p}{\cdot}{\bf q}}
&=& \bar\alpha_{\rm nl}  +
{\lambda + {4\over3}H\over 4}
 \int_0^\infty dy\,
K''(y)
\left(1+{\lambda y\over2}\right)^{n-1}
e^{-\chi y/\lambda\xi^2}
\nonumber\\
&=& \bar\alpha_{\rm nl} +
{\lambda + {4\over3}H\over 2} \int_0^\infty dy\,
f(y)
\left(1+{\lambda y\over2}\right)^{n-1}
e^{-\chi y/\lambda\xi^2},
\label{abG}
  \end{eqnarray}
which may be compared with KC, equation (3.36).
The important differences are the two extra factors in
the integrand, especially the exponential representing
the effect of intercommuting, and the fact that
$\lambda$ appears as well as $H$ in the factor multiplying
the integral.

It will be useful to define the dimensionless function
$F$ by
  \begin{equation}
2\bar\alpha_{\rm lin}  =
F\left({\zeta\over\xi},{\bar\xi\over\xi},
{t\over\xi}\right) =
{\lambda + {4\over3}H\over 2}
\int_0^\infty dy\, K''(y)
\left(1+{\lambda y\over2}\right)^{n-1}
e^{-\chi y/\lambda\xi^2}.
\label{Fdef}
  \end{equation}

If  $f = {1\over2}K''$  is assumed to have the
two-scale form (\ref{f2sc}), and if the length scales of
interest are large compared to $1/A$, then  $F$ is
expressible in terms of the incomplete gamma function:
 \begin{equation}
F \approx 2w\left(1 + {4H\over3\lambda}\right)
x^{-n} e^x \Gamma(n,x),
\label{FGam}
 \end{equation}
with
 \begin{equation}
x = {2B\over\lambda}+n.
\label{xdef}
 \end{equation}

Returning to the remaining terms in $\alpha$, we see
from (\ref{aldef1}) and (\ref{alG}) that  it may be
written in the form (\ref{alsplit}).
The function $\hat\alpha$ is given by
 \begin{eqnarray}
l\hat\alpha(l) &=& {(\lambda +
{4\over3}H)\over8K(l)} \int_0^l
dy_0\, [K'(y_0) + K'(l-y_0)] \nonumber\\
&&\times \int_0^\infty dy\,
[K'(y_0-y)+K'(l-y_0+y)]
\left(1+{\lambda y\over2}\right)^{n-1}
e^{-\chi y/\lambda\xi^2}.
\label{al1G}
 \end{eqnarray}

Unfortunately, it is no longer possible to perform
the $y_0$ integration explicitly without assuming the
form of $K$.

\subsection{Rates of change of length scales}

It is now easy to compute the rates of
change of the various scale lengths.

The rate of change of $L$ is given by (\ref{Ldstr})
(together with (\ref{abG})).  From (\ref{xid}), we thus
find
  \begin{equation}
{\dot\xi_{\rm str.}\over\xi}
= H\left({3\over2} - {\bar\alpha_{\rm nl}\over2} -
{1\over4}F\right).
\label{xidstr}
  \end{equation}

For the other length scales, we examine the
rate of change of $K$, given by (\ref{Kdstr}).
To do this, we need to examine the large- and small-$l$
behaviour of the parameters $\alpha$ and $\beta$.

First, we
consider the limit $l\to\infty$, where $K(l) \approx
2\bar\xi l + {\rm constant}$.  Substituting this form in the
expression (\ref{betaG}) for $\beta$, we find
 \begin{equation}
\beta(\infty) = \bar\alpha_{\rm nl} + {1\over2}G,
 \end{equation}
where the dimensionless function $G$ is given by
 \begin{equation}
G\left({\zeta\over\xi},{\bar\xi\over\xi},
{t\over\xi}\right) =
\left(\lambda+{2\over3}H\right)
\int_0^\infty
dy\,\left(1+{\lambda y\over2}\right)^{n-1}
e^{-\chi y/\lambda\xi^2}.
\label{Gdef}
 \end{equation}

This function can be explicitly evaluated in terms of
the incomplete gamma function, as
 \begin{equation}
G = 2\left(1+{2H\over3\lambda}\right)
n^{-n}e^n \Gamma(n,n).
\label{GGam}
 \end{equation}

If we use the same large-$l$ approximation in the
expression (\ref{al1G}) for $l\hat\alpha$ we see that the
$y_0$ integral is of order $\bar\xi^2 l$.  Clearly
therefore
 \begin{equation}
\hat\alpha(\infty) = 0.
 \end{equation}

These two results together yield
 \begin{equation}
{\dot{\bar\xi}_{\rm str.}\over\bar\xi}
= H[2\beta(\infty) - \bar\alpha] = H\left(\bar\alpha_{\rm nl} +
 G - {1\over2}F\right),
\label{xbdstr}
 \end{equation}

Next, we turn to the small-$l$ limit, using the
approximation $K(l) \approx l^2 - l^3/3\zeta$.  Now in
(\ref{betaG}), the expression in square brackets in the
integrand is clearly an even function of $l$, equal to
$l^2 K''(y) + {\cal O}(l^4)$.  We shall need the terms
in $\beta$ up to order $l$.  Since the denominator is
of order $l^2$, the $l^4$ term in the integrand is
irrelevant.  The integral that appears here is then
exactly the same as the one in (\ref{Fdef}).

It is no accident that $\beta(0)$ is almost exactly
equal to $\bar\alpha$.  In fact, it is physically obvious that
for short segments, the extension ${\bf r}$ and the length
$l$ must expand by identical factors.  There is an
apparent difference between the two, namely the
replacement of $(\lambda + {4\over3}H)$ by $(\lambda +
{2\over3}H)$.  However, this difference is spurious.  It
arose because in the one case we included the effect of
the rate of change of ${\bf p}(y_0)$, while in the other we
ignored the change of ${\bf r}$.  But for very short
segments, ${\bf r}$ and ${\bf p}$ have of course essentially
the same direction, and it is no longer reasonable to
neglect $\dot{\bf r}$.  Therefore in the small-$l$ limit,
we ought to replace  $(\lambda + {2\over3}H)$ by $(\lambda +
{4\over3}H)$.

In this way, we find
 \begin{equation}
\beta(l) \approx \beta_0 + \beta_1{l\over\zeta}
 = \bar\alpha_{\rm nl} + {1\over2}\left(1
+ {\l\over3\zeta}\right)F,
\qquad (l \ll\zeta).
\label{betasml}
 \end{equation}
The $l/\zeta$ term arises from expanding the denominator.

Next we turn to $\hat\alpha$.  For small $l$, the leading
term in the $y_0$ integral is
 \begin{equation}
\int_0^l dy_0\,[2\l][lK''(y)] = 2l^3 K''(y).
 \end{equation}
Again, therefore, we find the integral $F$ appearing:
 \begin{equation}
\hat\alpha(l) = \hat\alpha_0 + {\cal O}(l)
= {1\over2}F + {\cal O}(l),
\qquad (l\to 0).
\label{al1sml}
 \end{equation}
In this case, we shall not need the ${\cal O}(l)$ term.

To evaluate the $l^3$ term in (\ref{Kdstr}), we need to
examine the limiting behaviour of the function
$K_{(2)}(l)$.  As we saw in Section III F, as $l\to0$,
$K_{(2)} \propto l^5/\zeta$; to be specific let us assume
that
 \begin{equation}
K_{(2)} \approx C {l^5\over\zeta},\qquad l\to 0.
\label{Cdef}
 \end{equation}
Equivalently, $C$ is given by
 \begin{equation}
{K_{(2)}(l)\over K^2(l)} \approx {Cl\over\zeta},
\qquad l\to0.
 \end{equation}

Now from (\ref{Kdstr}), we find, using (\ref{betasml})
and (\ref{al1sml}),
 \begin{equation}
{\dot{\zeta}_{\rm str.}\over\zeta} = H(3\bar\alpha
- 2\beta_0 + 6\beta_1 - 12C\hat\alpha_0)
= H\left(\bar\alpha_{\rm nl}
 + {3-12C\over2}F\right),
\label{zedstr}
 \end{equation}

\section{GRAVITATIONAL RADIATION}

There is one final effect that we have not so
far considered, but which is in fact of great importance
in the long-term evolution of the string network:
gravitational back-reaction.  We have been able to
ignore it so far because it operates on a very
different length scale from most of the other effects.

Consider the gravitational radiation from a large
length $L$ of string.  Essentially the only scale that
can have any relevance here is the smallest
scale $\zeta$, which can roughly be identified with a
mean inter-kink distance.  We expect the rate of loss of
energy, or equivalently length, to be
\begin{equation}
\left(\partial L \over \partial t\right)_{\rm gr.rad.} =
- \Gamma G\mu {L\over\zeta},
\label{Ldgr}
\end{equation}
where $\Gamma$ is a constant of order $10^1$ or $10^2$.
In other words, the lifetime of the small-scale structure
would be of order $\zeta/\Gamma G\mu$.

It should be noted that numerically $\Gamma$ here may be
expected to differ somewhat from the values quoted in the
literature, which have mostly been derived
from studies of oscillating loops, because in those cases
the length scale used was the length of the loop, whereas
$\zeta$ is defined somewhat differently.  In fact, since
the typical loop size is probably a few times $\zeta$, our
$\Gamma$ is probably somewhat smaller.

Another way of expressing (\ref{Ldgr}) is to think of
the gravitational radiation as being generated by
each encounter between a pair of kinks, one left-moving
and one right-moving.  If we think of $\zeta$, roughly
speaking, as the mean inter-kink distance, we find
that the number of such encounters on a left-moving
segment of length
$L$ in a time interval $dt$ is
\begin{equation}
{2Ldt\over\zeta^2}.
\end{equation}
Hence (\ref{Ldgr}) is equivalent to saying that each
kink-kink encounter generates the release of an amount
of energy equal to
\begin{equation}
\case{1}/{2}\Gamma G\mu^2\zeta.
\end{equation}

The gravitational radiation from infinite strings has been
studied by several authors \cite{tv86,dg90,mh90}.  In
particular, Hindmarsh has obtained a formula
for the power emitted from encounters between left-moving
and right-moving sequences of small-angle kinks.  He finds
(ref. \cite{mh90}, eq. (23)) that
the power per unit length is
\begin{equation}
{dP\over dz} = 2\zeta(2)G\mu^2
\theta_u^2 \theta_v^2 \ln(d/r_K)d^{-1},
\label{mh23}
\end{equation}
where $\theta_u$ and $\theta_v$ are the kink angles of the
left- and right-moving kinks, $d$ is the inter-kink
distance and $r_K$ is the width of the string.  Since the
logarithm is slowly varying, it is reasonable to replace it
with a constant, of order $10$ to $10^2$.  The length
$d$ is of course related to our $\zeta$; by (\ref{zesmall}),
$d \sim \theta^2\zeta$.

Now let us consider how to apply this formula to our
problem, namely, how to estimate the rate of loss of
energy or length from a chosen
segment of left-moving string of length
$l$ and extension ${\bf r}$.  Consider first a relatively short
segment, containing a single small-angle kink of angle $2\theta$.
Clearly, we have
\begin{equation}
\theta^2 \sim {l^2 - {\bf r}^2\over l^2}.
\end{equation}
Hence the formula (\ref{mh23}) suggests that the rate of loss
of length should be proportional to $(l^2 - {\bf r}^2)/l^2$.

As in the case of stretching there is an important
consistency condition.  The long string can be divided up
conceptually into segments of any chosen length, and the
proportional rate of loss of length must be independent
of the choice, and must  agree with (\ref{Ldgr}), {\it i.e.}
 \begin{equation}
\bar{\dot l}_{\rm gr.rad.} = - {\Gamma G\mu\over\zeta}l.
\end{equation}

Putting these two requirements together, we find that the
expression for ${\dot l}_{\rm gr.rad.}$ must be
\begin{equation}
{\dot l}_{\rm gr.rad.} = - {\Gamma G\mu\over\zeta}\,l\,
{l^2 - {\bf r}^2\over l^2 - K(l)}.
\label{ldgr}
\end{equation}

So far, we have concentrated on the change in the length
$l$ of our segment, but we should also ask whether there
will be any change in the extension {\bf r}.  At first sight,
it might seem that the answer should be no.  Certainly
in the case where $l$ is large, gravitational radiation
will change the extension at most by an insignificant
amount.  However, it is also clear that for small $l$,
particularly if the segment is chosen to end near a kink,
there could be a significant reduction in ${\bf r}$.  Indeed,
as we shall see, it turns out that this is essential
for consistency.

If there is a reduction in ${\bf r}$ it seems reasonable to
suppose that, for segments of a given length $l$, it too
is proportional to $(l^2 - {\bf r}^2)$; certainly we would
expect it to vanish in the extreme case of a straight
segment, with ${\bf r}^2 = l^2$.  Let us therefore assume that
it takes the form
\begin{equation}
\dot{\bf r}_{\rm gr.rad.} = - {\Gamma G\mu \over\zeta}h(l)
{l^2 - {\bf r}^2\over l^2 - K(l)}{\bf r},
\label{rdgr}
\end{equation}
where $h(l)$ is an as yet unknown function.

It also seems plausible to assume that for very large
segments gravitational radiation has no significant
effect on the overall extension.  This would imply that
\begin{equation}
h(l)\to 0 \;{\rm as} \; l\to\infty.
\label{hinf}
\end{equation}

We are now in a position to evaluate the change due
to emission of gravitational radiation in the probability
distribution $p[{\bf r}(l)]$.  It is obtained in exactly
the same way as in the case of stretching, from the
analogue of (\ref{pdstr}), namely
\begin{equation}
\left(\partial p\over\partial t\right)_{\rm gr.rad.} =
- {\partial\over\partial l}({\dot l}_{\rm gr.rad.} p)
- {\partial\over\partial{\bf r}}{\cdot}(\dot{\bf r}_{\rm gr.rad.} p)
+ {\dot L_{\rm gr.rad.}\over L}p.
\label{pdgr}
\end{equation}

 From this, we easily find
\begin{equation}
\left(\partial K\over\partial t\right)_{\rm gr.rad.} =
- {\partial\over\partial l} \left(
\overline{{\dot l}_{\rm gr.rad.} {\bf r}^2}\right)
+ 2\overline{{\bf r}{\cdot}\dot{\bf r}_{\rm gr.rad.}}
+ {\dot L_{\rm gr.rad.}\over L} K.
\end{equation}
Substituting from (\ref{Ldgr}), (\ref{ldgr}) and (\ref{rdgr}),
and using the identity
\begin{equation}
\overline{{\bf r}^2{l^2 - {\bf r}^2\over l^2 - K}}
= \overline{{\bf r}^2} - \overline
{{\bf r}^2{{\bf r}^2 - K\over l^2 - K}} =
K - {K_{(2)}\over l^2 - K},
\end{equation}
we obtain
\begin{equation}
\left(\partial K\over\partial t\right)_{\rm gr.rad.} =
{\Gamma G\mu\over\zeta} \left\{lK' -
{\partial\over\partial l}\left({lK_{(2)}\over l^2-K}\right)
- 2hK + 2h{K_{(2)}\over l^2-K}\right\}.
\label{Kdgr}
\end{equation}

As before, we can now find the rates of change of the
various length scales.  From (\ref{xid}) and
(\ref{Ldgr}), we get
\begin{equation}
{\dot\xi_{\rm gr.rad.}\over\xi} =
{\Gamma G\mu\over2\zeta}.
\label{xidgr}
\end{equation}

Similarly, from (\ref{xbd}) and (\ref{Kdgr}), we find
\begin{equation}
{\dot{\bar\xi}_{\rm gr.rad.}\over\bar\xi} =
{\Gamma G\mu\over\zeta}.
\label{xbdgr}
\end{equation}

When we come to examine the third length scale, $\zeta$,
we can see why the presence of $h$ is essential
for consistency.  Consider the limit $l\to0$.  For
small $l$, the leading terms in $K(l)$ are given by
(\ref{Ksmall}), so $l^2 - K \approx l^3/3\zeta$.
Moreover, as we saw earlier,
$K_{(2)} \approx C l^5/\zeta$.
Thus we find that the expression in the large braces
in (\ref{Kdgr}) has a term that behaves like $l^2$
for small $l$. This is inconsistent with the assumed form of
$K$.  Hence for consistency of our approximation
we must assume that the
coefficient of $l^2$ vanishes, which requires
that
 \begin{equation}
h(0) = {1 - {9\over2}C \over 1 - 3C}.
\label{h0}
 \end{equation}
The value of $h'(0)$ turns out also to be important,
but is not constrained by any consistency requirement.

Then, by (\ref{zed}), the $l^3$ term in (\ref{Kdgr})
yields
 \begin{equation}
{\dot{\zeta}_{\rm gr.rad.}\over\zeta} =
 \hat C{\Gamma G\mu\over\zeta}.
\label{zedgr}
 \end{equation}
where $\hat C$ is another constant, related in a somewhat
complicated way to the leading terms in the power-series
expansions of $K$, $K_{(2)}$ and $h$.  Specifically, if
 \begin{eqnarray}
K(l) &=& l^2 - {l^3\over 3\zeta}
+ {k_4 l^4\over12 \zeta^2} + \dots,
\nonumber\\
K_{(2)}(l) &=& {C l^5\over\zeta} - {C_6 l^6\over\zeta^2} + \dots,
\\
h(l) &=& h(0) - {h_1l\over\zeta} + \dots,
\nonumber
 \end{eqnarray}
then
 \begin{equation}
\hat C = -3(1 - 12C_6 + 3k_4 C)
+ 2h(0)(1 - 9C_6 + \case{9}/{4}k_4 C)
+ 6 h_1(1 - 3C).
\label{Chatdef}
 \end{equation}

 From (\ref{K2l0G}) we see that if the
model of Gaussian-distributed small-angle kinks is
correct, then
 \begin{equation}
C = {\overline{\theta^2}\over 15}.
\label{Capp}
 \end{equation}
In any event, it seems likely that $C\ll1$.

As a good first approximation, we may set $C=0$ and
$C_6 = 0$,
which means that $h(0) = 1$ and the
value of $k_4$ becomes irrelevant.  Then
(\ref{zedgr}) becomes
\begin{equation}
{\dot{\zeta}_{\rm gr.rad.}\over\zeta} =
 (6h_1 - 1){\Gamma G\mu\over\zeta}.
\label{zedgr2}
\end{equation}
Note, however, that the value of $h_1$ is clearly
important.  The sign of this term is crucial, as we shall
see, in determining the nature of the solution.  In our
rough approximation, $\hat C > 0$ requires $h_1 >
{1\over6}$.

\section{OVERALL EVOLUTION EQUATIONS}

We can now combine all the various terms together to give
composite rate equations for $L$, for $p[{\bf r}(l)]$, and
for $K(l)$.  We hope to return to these equations
at a later date.  For the moment, however, we shall not
write them down  explicitly, but concentrate instead on
the equations for the three length scales.

\subsection{Rates of change of length scales}

We begin with $\xi$.  From (\ref{xidstr}),
(\ref{xidllf}) and (\ref{xidgr}), we obtain
 \begin{equation}
{\dot\xi\over\xi} = H\left({3\over2} -
{\bar\alpha_{\rm nl}\over2} -
{1\over4}F\right)
+ {1\over2\bar\xi}E
+ {\Gamma G\mu\over2\zeta},
\label{xid1}
 \end{equation}
where the dimensionless functions $F$ and $E$ were
defined in (\ref{Fdef}) and (\ref{Edef}).

For the other large length scale $\bar\xi$, we have,
from  (\ref{xbdstr}), (\ref{xbdlsi}), (\ref{xbdllf})
and  (\ref{xbdgr}),
 \begin{equation}
{\dot{\bar\xi}\over\bar\xi} = H\left(\bar\alpha_{\rm nl} +
G - {1\over2}F\right)
- {\chi\over w}{\bar\xi\over\xi^2}
+ {1\over2\bar\xi} I
+ {\Gamma G\mu\over\zeta},
\label{xbd1}
 \end{equation}
where $I$ was defined in (\ref{Idef}).

Finally, we turn to the small-distance scale length
$\zeta$. Using (\ref{zedstr}),
(\ref{zedlsi}), (\ref{zedllf}) and (\ref{zedgr}), we
find
 \begin{equation}
{\dot{\zeta}\over\zeta} = H\left(\bar\alpha_{\rm nl} +
{3-12C\over2}F\right)
- {\chi\zeta\over\xi^2}
+ {k\over\bar\xi}E
+ {\Gamma G\mu\over\zeta} \hat C,
\label{zed1}
 \end{equation}
where $\hat C$ was defined in (\ref{Chatdef}).

\subsection{Estimation of parameters}

To determine the outcome of the evolutionary process,
we first have to estimate how the unknown functions $E,
F, G, I$ depend on the ratios of length scales, and the
likely magnitude of the additional parameters
$\bar\alpha_{\rm nl}, w, C, \hat C$ and $k$.

In some ways the most basic parameter is $E$, which
governs the rate of loop formation $\lambda$ via the
relation $\lambda = E/\bar\xi$.  As we argued
in Section VIB, if initially all the length scales are
of comparable magnitude, then $\lambda$ will be very small
because almost all loops formed reconnect.  In that
situation, $E\ll 1$.  Only when the upper cutoff
$\xi^2/t$ has grown to be significantly larger than
$\zeta$ do many loops start to survive.  Once that
happens, we expect $E$ to become of order unity.  At
present, we are not able to calculate the value very
precisely, because it is strongly influenced by the
intermediate-scale angular correlation effect discussed
in Section VIB, for which we have only a very
qualitative treatment.  Fortunately, the precise value
of $E$ is not critical, because it appears as a common
factor in several of the important terms.

We note that, according to (\ref{abnldef}),
 \begin{equation}
\bar\alpha_{\rm nl} \approx {\lambda\zeta\over2}
= {E\zeta\over2\bar\xi}.
 \end{equation}
This parameter will be very small,
$\bar\alpha_{\rm nl} \ll 1$, throughout most
of the relevant region of parameter space, initially
because $E\ll 1$ and later because $\zeta\ll\bar\xi$.
There might be just a short period in which it becomes
non-negligible, when $\xi$ has grown large compared to
$\bar\xi$, but $\bar\xi$ and $\zeta$ are still comparable
in magnitude; but for the most part it can be safely
neglected.

Now let us turn to the function $G$ which, in
(\ref{GGam}), was expressed in
terms of the incomplete gamma function.  It depends
primarily on the value of $\lambda\xi$.  According to
(\ref{ndef}), when $\lambda\xi$ is small, $n\gg1$.  Then we
can use the fact that asymptotically $\Gamma(n,n) \approx
{1\over2}\Gamma(n)$ to get
 \begin{equation}
G \approx \left(1 + {2H\over
3\lambda}\right) \sqrt{2\pi\over n}
 \approx \sqrt{\pi\over\chi}(\lambda +
\case{2}{3} H)\xi
\qquad ({\rm small}\;\lambda\xi).
 \end{equation}
In the opposite limit of large $\lambda\xi$, where $H\xi$
is certainly negligible, we have
 \begin{equation}
G \approx -2{\rm Ei}(-n)
 \approx -2\ln n \approx 4\ln(\lambda\xi)
\qquad ({\rm large}\;\lambda\xi).
 \end{equation}

Next, let us consider the function $F$,
given by (\ref{Fdef}). There are two large scales
involved here (quite apart from the small scale
$\zeta$).   The function $F$ depends  primarily on the two
variables $\lambda\xi$ and  $\lambda\bar\xi$, though it also has a
weak  dependence on the small length scale.  Note
that, according to (\ref{xdef}), the argument $x$ of
the incomplete gamma function in (\ref{FGam}) is
 \begin{equation}
x = {2B\over\lambda} + n =
{2w\over\lambda\bar\xi} + {2\chi\over\lambda^2\xi^2}.
 \end{equation}
Consider first the case where $\lambda\bar\xi \gg
(\lambda\xi)^2$.  Then $x\approx n$ and so
(if $H\ll\lambda$)
 \begin{equation}
F\approx wG, \qquad \big(\lambda\bar\xi \gg
(\lambda\xi)^2\big).
 \end{equation}
Note that, directly from the integral definitions
(\ref{Fdef}) and (\ref{Gdef}), it follows that, at least
when $H$ is negligible, we always have $F<G$.

In the opposite limit, where $\lambda\bar\xi \ll
(\lambda\xi)^2$, we have $x \approx 2w/\lambda\bar\xi
\gg n$, and consequently $F \ll G$.  If $\lambda\bar\xi \ll 1$,
we can use the large-$x$ form of the incomplete gamma
function to get
 \begin{equation}
F \approx (\lambda + \case{4}{3} H)\bar\xi =
\left(1 + {4H\over3\lambda}\right)E
\qquad ({\rm small}\;\lambda\bar\xi).
 \end{equation}
On the other hand, if $\lambda\bar\xi$ is large, we can use
the small-$x$ approximation, obtaining
  \begin{equation}
F \approx 2w \ln(\lambda\bar\xi),
\qquad \big((\lambda\xi)^2 \gg
\lambda\bar\xi \gg 1).
 \end{equation}

In the intermediate region, where $\lambda\bar\xi$ and
$\lambda\xi$ are of order unity, $F$ and $G$ are both
likely to be, very roughly, of order unity, with $F <
G$.  Strictly speaking, the expressions for $F$ and $G$
are both based on a linear approximation and cease to be
valid for large $\lambda\xi$ or $\lambda\bar\xi$.  Recall that
$F/2$ is the linear contribution to the value of $\bar\alpha$.
 From the simulations, we know that $\bar\alpha \sim 0.2$, so
$F$ is almost certainly significantly less than unity.

Now we turn to the function $I$ defined by
(\ref{Idef}).  Initially, if all the length scales are
comparable, $I$, like the other functions, will be
small, because the loop formation probability is small.
Consider, however, a later time at which loop formation
has started and $\bar\xi$ has grown to be $\gg\zeta$.
The integral is then dominated by the region where
$l_1$ is a few times $\zeta$.  In that region,
$K_1\ll4\bar\xi l_1$.  Hence we have to a good
approximation,
 \begin{equation}
I \approx 2\bar\xi \int_0^\infty dl\,l\Lambda(l)
{K(l)-Q(l)\over K(l)+Q(l)}.
 \end{equation}
Clearly, the ratio $Q/K$ plays a very important role
here.  If it is approximately constant over the
relevant range, then we get a very simple expression
for $I$:
 \begin{equation}
I \approx 2 {K-Q\over K+Q} \lambda\bar\xi
= 2{K-Q\over K+Q} E.
\label{Iapp}
 \end{equation}

Now, how large is $Q$ likely to be?  If the original
expressions (\ref{Th1}) and (\ref{delta0def}) for the
loop-formation probability function $\Theta({\bf r},l)$
were  correct, we should expect
 \begin{equation}
{1\over Q} = {1\over K} + {2a\over3l^2},
\label{Qapp}
 \end{equation}
where the second term arises from the volume factor
$\Delta$.  But in addition to this, we saw in Section
VIB that the angular correlation effect would be
expected to contribute another term to $1/Q$.  Although
we have at present no means of estimating this
contribution with any precision, it seems reasonable to
expect that it too would be proportional to $1/l^2$ and
of a similar order of magnitude.  Even from
(\ref{Qapp}), we see that $1/Q$ should be significantly
larger than $1/K$.  The additional angular correlation
term enhances this effect, suggesting that $1/Q$ should
be very substantially larger.  Consequently, $Q\ll K$
and it may be a reasonable first approximation to set
$Q/K=0$, which would mean that $I\approx 2E$.  In any
event, we expect that the ratio $I/E$ will be not too
far below 2.  We shall see later that $I/E$ is closely
related to the parameter $q$ introduced in CKA
\cite{CKA}; in fact $I/E\approx q-1$.

There are several other unknown parameters that enter
our evolution equations.  The parameter $w$ is defined
in terms of the large-$l$ behaviour of the variance
function $K$: for large $l$, $K\approx 2\bar\xi l -
2\bar\xi^2/w$.  It was originally introduced in terms of
the illustrative two-exponential model of Section IIIA,
which suggests that it is limited to the range
$0<w<1$; a typical value might be $1\over2$.  So long
as it is not {\it very\/} small compared to 1, the precise
value of $w$ is not critical.

The parameter $C$ is defined by (\ref{Cdef}) in terms
of the small-$l$ behaviour of the function
$K_{(2)}(l)$.  Simple models of kinks suggest, {\it
e.g.}\ as in (\ref{Capp}), that $C$ is small compared to
unity, say of order 0.1 or less.  Within this general
range, the precise value is probably not significant.

The effect of gravitational radiation on the small
length scale $\zeta$ is governed, according to
(\ref{zedgr}), by the parameter $\hat C$, defined by the
rather complicated relation (\ref{Chatdef}).  In view
of the number of independent parameters that contribute
to this relation, it seems to be very difficult to
estimate $\hat C$ from first principles.  However, on
physical grounds it seems clear that $\hat C$ should be
positive, and presumably of order unity.  The effect of
gravitational back-reaction {\it must\/} be to smooth
out the small-scale kinks on the string, and the
expected time-scale for this process must be
of order $\zeta/\Gamma G\mu$.

Finally, we have to consider the parameter $k$.  Like
$I/E$, $k$ is related to the parameter $q-1$ defined in
CKA \cite{CKA}, except that it is not concerned with
the large-scale structure described by $\bar\xi$ but with
the small-scale kinkiness described by $\zeta$.

One indication of its magnitude can be obtained by
considering a related but slightly different
parameter.  Consider loops of length $l$ and extension
${\bf r}$.  The probability of finding such a loop is
proportional to $p[{\bf r}(l)]\Theta({\bf r},l)$.  Hence if we
choose a loop of length $l$ at random, the variance of
its extension is
 \begin{equation}
\left({1\over K} + {1\over Q}\right)^{-1}
= {KQ\over K+Q}.
 \end{equation}
Since we know that $Q$ is small compared to $K$, this
variance is $\ll K$.  This statement applies to the
whole loop, not to a segment on the loop, but of course
the two are related.  If a loop typically has an
extension much less than that of a similar piece of
long string, the same must be true for a small segment
on the loop, though the effect is probably less
dramatic.  So we must expect
$\overline{{\bf r}^2_{\rm loop}} < K(l)$,
which implies $k > 0$.  It is not so easy to estimate
its magnitude, but as we shall see there are reasons for
believing that it cannot be large.

\subsection{Equations for scaling variables}

In our previous work \cite{CKA}, to discuss the
possibility of scaling, we expressed both our length
scales as fractions of the horizon distance $R\tau$.  It
turns out, however, that we can obtain slightly simpler
equations if instead of the horizon we use the
expansion time, $1/H$.  We define the three
dimensionless ratios
 \begin{equation}
\gamma = {1\over H\xi},\quad
\bar\gamma = {1\over H\bar\xi},\quad
\epsilon = {1\over H\zeta}.
 \end{equation}
In the radiation era, $\gamma$ and $\bar\gamma$ are identical to
the variables used in KC and CKA, but in the
matter-dominated era they are half as large.

We also define $p$ so that $R\propto t^{1/p}$, {\it i.e.},
$H=1/pt$.  Thus $p=2$ in the radiation era and
$p={3\over2}$ in the matter era.

The dimensionless functions $F,G,E,I$ may now be
regarded as functions of the ratios of these variables,
{\it e.g.},
 \begin{equation}
E = E\left(
{\gamma\over\epsilon},{\gamma\over\bar\gamma},
{\gamma\over p}\right).
 \end{equation}

Then, substituting into the evolution equations and
dividing by $H$, we obtain
 \begin{eqnarray}
-pt{\dot\gamma\over\gamma} &=& -p +
\left({3\over2}
-{\bar\alpha_{\rm nl}\over2} -
{F\over4}\right)
+{E\over 2}\bar\gamma
+ {\Gamma G\mu\over2}\epsilon,
 \nonumber
\\
-pt{\dot{\bar\gamma}\over\bar\gamma} &=& -p+
\left(\bar\alpha_{\rm nl} +
G - {F\over2}\right)
- {\chi\gamma^2\over w\bar\gamma}
+ {I\over2}\bar\gamma
+ \Gamma G\mu\epsilon,
 \label{scald}
\\
-pt{\dot{\epsilon}\over\epsilon} &=& -p+
\left(\bar\alpha_{\rm nl}  +
{3-12C\over2}F\right)
-{\chi\gamma^2\over\epsilon}
+kE\bar\gamma
+\Gamma G\mu\hat C\epsilon.
\nonumber
 \end{eqnarray}

It is trivial to write down equations for the rates of
change of the {\it ratios\/} of length scales.  In
particular, we find
 \begin{equation}
p{d\ln(\bar\xi/\zeta)\over d\ln t}
= G - 2(1-3C)F - \chi\left(
{\gamma^2\over w\bar\gamma} - {\gamma^2\over\epsilon}
\right) + \left({I\over2} - kE\right)\bar\gamma
- \Gamma G\mu(\hat C - 1)\epsilon.
\label{xbzed}
 \end{equation}
This will be useful later.

\subsection{Scaling of $\xi$ and $\bar\xi$}

We now return to the main issue: what do the evolution
equations tell us about how the length scales evolve?

Let us suppose, as in Section VIB, that we
start with all three length scales comparable in
magnitude, and at least somewhat smaller than the
horizon size.  Initially, most loops that form will
reconnect, so $E$ will be small, as will the other
dimensionless functions $F$, $G$ and $I$.  It is then
clear that $\xi$ will start to grow, because of the
stretching effect, while initially $\bar\xi$ and $\zeta$ will
not.  This will continue until loop production starts to
be significant, when the upper cutoff, $\xi^2/t$ exceeds
a few times $\zeta$.  In that region, we have
$\zeta\sim\bar\xi\ll\xi$.

As $\lambda$ grows, it will first reach the point where
$\lambda\xi \sim 1$, while $\lambda\bar\xi$
is still small.  In that region, $E$, $F$ and $I$
remain small, but $G$ becomes of order unity.  From
(\ref{xbzed}) it is clear that the ratio $\bar\xi/\zeta$
will then start to grow.

Some time later, we may reach the point where
$\lambda\bar\xi \sim 1$ and $\lambda\xi \gg 1$.  If so,
$E$, $F$ and $I$ would be of order unity but $G$ would
be even larger, so $\bar\xi$ would grow rapidly
until it caught up with $\xi$.

Assuming the gravitational radiation terms are still
negligible, do we then reach a regime where $\xi$ and
$\bar\xi$ at least approximately scale?

For this to happen, the right hand sides of the first
two equations in (\ref{scald}) must vanish.  If we
suppose for a moment that the values of $E,F,G,I$ are
known, then we can solve for $\gamma$ and $\bar\gamma$.  The first
equation always yields a positive value of $\bar\gamma$:
 \begin{equation}
\bar\gamma = {1\over E}\left(2p - 3 + \bar\alpha_{\rm nl}
+ {F\over2}\right).
\label{gbsc1}
 \end{equation}
The condition that the second equation yields a
positive value of $\gamma^2/\bar\gamma$ is
 \begin{equation}
(-2p +2\bar\alpha_{\rm nl} + 2G - F) +
{I\over E}\left(2p-3 + \bar\alpha_{\rm nl}
+ {F\over 2}\right) > 0.
\label{scpos}
 \end{equation}
The ratio $I/E$ plays the same role as $q-1$ in CKA.
Scaling requires a sufficiently large value of this
parameter.

If the condition (\ref{scpos}) is {\it not\/}
satisfied, scaling cannot be achieved at the given
values of $E,F,G,I$.  But of course these are not
constants.  If $\bar\gamma$ satisfies (\ref{gbsc1}) but
(\ref{scpos}) is violated, then clearly $\bar\gamma$ will
start to grow.  The right hand side of the first of
equations (\ref{scald}) then becomes positive, so $\gamma$
begins to fall.  In other words, $\xi$ grows faster
than $\bar\xi$.  This of course affects the values of the
various functions.  If $E=\lambda\bar\xi$ remains of order
unity, then $\lambda\xi$ must grow, leading eventually to
an increase in $G$ (and a smaller increase in $F$).  This
in turn decreases the right hand side of the first
equation in (\ref{scald}) and increases that of the
second, leading us back towards scaling.  It seems likely
therefore that the parameters will adjust to reach the
point where $\xi$ and $\bar\xi$ do indeed scale.

We must also consider what happens to the third length
scale in this partial scaling regime.  To find out, we
examine the third of the equations (\ref{scald}).

Let us assume that $\gamma, \bar\gamma \ll \epsilon
\ll (\Gamma G\mu)^{-1}$.  Then both the intercommuting and
gravitational radiation terms in the third equation are
negligible.  The essential question is: what is the
sign of the right hand side?  Using the partial-scaling
value of $\bar\gamma$, given by (\ref{gbsc1}), we find that
for $\epsilon$ to grow we need
 \begin{equation}
k < {p - \bar\alpha_{\rm nl} - {3\over2}(1-4C)F
\over 2p - 3 + \bar\alpha_{\rm nl} + {1\over2}F}.
\label{klim}
 \end{equation}
For example, if we take $\bar\alpha_{\rm nl} = 0$, $C=0$,
$F=0.4$ and $p=2$, then we require $k < {7\over6}$.
(For a similar value of $F$, the condition is less
restrictive in the matter-dominated era, where $p =
{3\over2}$.)

In passing, it is worth asking what would happen if the
condition (\ref{klim}) were violated.  In that case of
course $\epsilon$ would decrease; $\zeta$ would start to catch
up with $\xi$ and $\bar\xi$.  If the dimensionless
functions $F,G,E,I$ were simply constants, $\zeta$
would continue to grow faster than $\xi$ and $\bar\xi$
until it actually exceeded them.  This is obvious
nonsense.  What really happens is that eventually
$\lambda$ starts to drop, thus returning us to a situation
closer to our starting point.  In fact, it appears that
we would then reach a complete scaling regime even
without invoking the gravitational radiation term.

This should not be a surprising conclusion.  We have
already noted that $k$, like $I/E$, is analogous to the
parameter $q-1$ of our earlier work; $k$ represents the
fractional excess small-scale kinkiness of a loop as
compared to a long string.  If $k$ is large, this means
that the loop-formation process very efficiently
removes small-scale kinkiness, so that complete scaling
of all three length scales is indeed possible.

What the simulations suggest, however, is a very
different scenario, in which $\xi$ and $\bar\xi$ at least
approximately scale, but $\zeta$ does not.  This is
exactly what we expect if the inequality (\ref{klim})
{\it is\/} satisfied.

Our conclusion here is really a straightforward
generalization of our earlier results.  In CKA
\cite{CKA}, we showed that scaling should occur if the
parameter $q-1$ exceeds a critical value.  In our present
work, we treat separately the large-scale and
small-scale kinkiness.  The parameter that represents
the excess large-scale kinkiness is $I/E$.  Provided
it is big enough, $\xi$ and $\bar\xi$ will scale (and, as
we have argued, that is not a restrictive condition,
because the parameters can adjust themselves until it
is fulfilled).  On the other hand, for $\zeta$ to scale
as well, without $\bar\xi/\zeta$ becoming large,
the excess small-scale kinkiness parameter $k$ would
have to exceed a critical value.  This condition is
almost certainly {\it not\/} fulfilled in the real
system.

\subsection{Possibility of complete scaling}

Let us assume then that (\ref{klim}) holds.  Then while
$\gamma$ and $\bar\gamma$ settle down to scaling values,
$\epsilon$ continues to grow.

This {\it transient scaling\/} regime continues until
$\epsilon$ grows to be of order $(\Gamma G \mu)^{-1}$,
so that the gravitational radiation terms become
important.  The question is: do we then reach a new
scaling regime in which all three length scales grow in
proportion to $t$?  Equivalently, do we reach a fixed
point of the three equations (\ref{scald})?

If there is a complete scaling solution, it is in the
region where $\epsilon\gg\gamma\sim\bar\gamma$.
This means that the intercommuting term in the third
equation, $\chi\gamma^2/\epsilon$, is very small.
If we neglect it,  then it is straightforward, for given
values of $E,F,G,I$, to solve the first and third
equations for $\bar\gamma$ and $\epsilon$, and then to
find $\gamma$ from  the second.

The solutions for $\bar\gamma$ and $\epsilon$ are given by
 \begin{eqnarray}
\bar\gamma &=& {(2p-3)\hat C - p + (\hat C + 1)\bar\alpha_{\rm nl}
+{1\over2}(\hat C+3-12C)F \over (\hat C-k)E},
\nonumber
\\
&&
\label{gbepsc}
\\
\epsilon &=& {p - (2p-3)k - (1+k)\bar\alpha_{\rm nl}
-{1\over2}(3-12C+k)F \over (\hat C-k)\Gamma G\mu}.
\nonumber
 \end{eqnarray}

A consistent solution requires that the numerators of
both these expressions be positive.  This places a
lower bound on $\hat C$, namely
 \begin{equation}
\hat C > {p - \bar\alpha_{\rm nl} - {3\over2}(1-4C)F
\over 2p - 3 + \bar\alpha_{\rm nl} + {1\over2}F}
> k,
\label{Chlim}
 \end{equation}
where the second inequality is merely (\ref{klim}).

We recall that, according to (\ref{zedgr}), $\hat C$ is the
parameter that defines the efficiency with which
gravitational radiation removes small-scale kinkiness.
It is not at all surprising to find that this parameter
must exceed some critical value if scaling is to be
achieved.  In fact, the second inequality in
(\ref{Chlim}) ensures that loop formation is {\it
not\/} sufficient to induce scaling of $\zeta$, while
the first inequality ensures that gravitational
radiation {\it is\/} sufficient.

The corresponding scaling value of $\gamma$ is found from
the second of equations (\ref{scald}).  As before,
there is a consistency condition stemming from the
requirement that $\gamma^2/\bar\gamma$ be positive.
This condition may be written
 \begin{equation}
(p - 3 + 2\bar\alpha_{\rm nl} + G)(\hat C - k)
> \left(1 - {I\over2E}\right)
\{[2p - 3 +\bar\alpha_{\rm nl} + \case{1}{2}F]\hat C
- [p - \bar\alpha_{\rm nl} - \case{3}{2}(1-4C)F]\}.
\end{equation}
Since both factors on the right hand side are
definitely positive (the second in virtue of
(\ref{Chlim})), this requires a minimum value of $G$:
 \begin{equation}
G + 2\bar\alpha_{\rm nl} > 3 - p.
\label{Gmin}
 \end{equation}
As in the earlier transient scaling regime, if this
condition is not satisfied, then $\bar\gamma$ will start to
grow and $\gamma$ to fall until it is.

We conclude that the parameters $\hat C$ and $k$ are the
most important in determining whether complete scaling
will be achieved.  The essential condition is
(\ref{Chlim}).

\subsection{Stability}

Let us assume that a scaling solution exists and ask
whether it is stable.
To study this question, we consider small perturbations
in the three scaling parameters, $\delta\gamma$,
$\delta\bar\gamma$ and $\delta\epsilon$, looking for solutions
proportional to $t^\lambda$.  Strictly speaking, of course,
the functions $E,F,G,I$ are functions of these
parameters, but we shall assume that they vary slowly
over the relevant range, so that they may be treated
for the purpose of the stability analysis as constants.

The condition for
stability is that the three eigenvalues $\lambda$
are all negative.  In other words, if we linearize the
set of equations (\ref{scald}) about the presumed
scaling solution, the resulting $3\times3$ matrix must
be positive definite.  This matrix is
 \begin{equation}
\left[
\begin{array}{ccc}
0 & {E\over2} & {\Gamma G\mu\over2} \\
-{2\chi\gamma\over w\bar\gamma} &
{\chi\gamma^2\over w\bar\gamma^2} + {I\over2} &
\Gamma G\mu \\
- {2\chi\gamma\over\epsilon} & kE &
{\chi\gamma^2\over\epsilon^2} + \Gamma G\mu\hat C
\end{array}
\right].
 \end{equation}
There are three required conditions, the positivity of
the trace, of the sum of $2\times2$ principal minor
determinants and of the determinant.  The first is
automatic.  The other two are
 \begin{equation}
{\chi\gamma\over w\bar\gamma} E + \Gamma G\mu
\left[\hat C\left({\chi\gamma^2\over w\bar\gamma^2}
+ {I\over2}\right) - kE\right]
+ \chi\left[{\gamma^2\over\epsilon^2}\left(
{\chi\gamma^2\over w\bar\gamma^2} + {I\over2}
\right) + {\gamma\over\epsilon}\Gamma G\mu\right]
> 0,
 \end{equation}
and
 \begin{equation}
(\hat C-k)\Gamma G\mu{E\gamma\over w\bar\gamma} +
{\gamma\over\epsilon}\left[\Gamma G\mu\left(
{\chi\gamma^2\over w\bar\gamma^2} + {I\over2} -
E\right) + {E\chi\gamma^2\over w\bar\gamma\epsilon}
\right] > 0.
 \end{equation}
In both cases, the terms are ordered according to the
number of powers of the small quantities
$\Gamma G\mu$ and $\gamma/\epsilon$ that they contain.
We see that in both equations, the leading terms are
automatically positive.  Thus we conclude that if the
scaling solution exists it is almost certainly stable.
The essential conditions are the bounds (\ref{Chlim}) on
$k$ and $\hat C$ required for the solution to exist.

\section{CONCLUSIONS}

We have analysed in considerable detail the various
processes that affect the evolution of a network of
cosmic strings --- stretching, intercommuting, loop
formation and gravitational radiation.  We have derived
a set of coupled evolution equations for the three
length scales that describe the configuration ---
$\xi$, the inter-string distance, $\bar\xi$, the
large-scale persistence length along the string, and
$\zeta$, which characterizes the small-scale kinkiness.

We showed that under reasonable assumptions $\xi$ and
$\bar\xi$ reach  a scaling regime, growing in
proportion to the horizon, while $\zeta$ grows slowly if
at all.  This continues until the ratio $\xi/\zeta$
becomes large enough for gravitational radiation
effects to be significant.  Thereafter, it is likely
that a new scaling regime is reached, in which all
three lengths scale, with $\xi\sim\bar\xi$ and
$\zeta/\xi\sim \Gamma G\mu$.

These conclusions depend on estimates of various
functions and parameters that are still somewhat
unreliable.  The most significant parameters are $k$,
which describes the excess small-scale kinkiness of
loops as compared with long pieces of string, and
$\hat C$, which determines the rate at which
gravitational back-reaction smoothes the small-scale
kinkiness.  The fact that small-scale structure is seen
to build up in the simulations strongly suggests that
loop formation alone is not able to smooth the
small-scale kinkiness, or in other words that $k$ is
less than the critical value, (\ref{klim}).  The
essential condition for a full scaling solution to be
reached is that $\hat C$ exceeds the critical
value, {\it i.e.}, that gravitational radiation {\it is\/}
effective in smoothing the small-scale kinks.

It would obviously be desirable to be able to estimate
the parameters $k$ and $\hat C$ from first principles.  We
hope to do this at a later date.

Another key part of the discussion concerns the
build-up of angular correlations between the left- and
right-movers due primarily to loop formation.  We have
given a fairly precise account of the small-scale
effect, leading to correlations between the ${\bf p}$
and ${\bf q}$ vectors, but the treatment of the
intermediate-scale correlations involving segments of
length a few times $\zeta$ is still rather sketchy.  It
is therefore very important to try to improve this part
of the discussion.  We hope to return to this question
shortly.

Once some of the remaining uncertainties have been
resolved, we should be in a position to give a more
accurate description of the scaling regime.  It is very
important to realise that the simulations performed so
far, which neglect gravitational radiation, may have
given misleading information about the typical scales
involved.  From the evolution equations,
(\ref{scald}), it is clear that when the gravitational
radiation terms come into play, the effect will be to
make $\bar\gamma$ decrease.  (It is easy to verify that the
value of $\bar\gamma$ in (\ref{gbepsc}) is less than
(\ref{gbsc1}) provided that (\ref{klim}) is satisfied.)
In other words, the typical persistence length in the
final scaling regime is probably a larger fraction of the
horizon distance than it was in the temporary scaling
regime where gravitational radiation was negligible ---
the only one so far accessible to simulations.  Whether
$\gamma$ also decreases is less obvious.  That seems to
depend primarily on the magnitude of the ratio $I/E$.  We
hope to provide a more detailed analysis of this behaviour
in a future publication.

There are many interesting questions that we should be
able to address. With an analytic model it should be
possible to estimate the typical density fluctuations, and
hence microwave anisotropies, generated by a network of
strings. Such an approach has previously been adopted in
\cite{tk85} using the one-scale model, but it is essential
that we understand the role the small-scale structure
plays in determining the form and amplitude of the
fluctuations. For example, do we see any evidence of the
effects from the string tension renormalisation as
indicated in the models of Albrecht and Stebbins
\cite{AS92}?

Another interesting aspect which we hope to address
concerns the transition period from radiation to matter
domination. What happens to the scaling solutions in the
two regimes; do they smoothly evolve from one scaling
value to the other? A more detailed analysis of the
evolution equations will enable a determination to be made
of the approach to scaling. We have established the
conditions required for scaling but have not discussed the
time scales over which such conditions could be met. In
particular, it would be interesting to examine the
duration of the transient scaling regime and the way in
which it is goes over to full scaling.  This is important
because the numerical simulations only directly give us
information about evolution over a few expansion times.
It might be possible to tackle this problem numerically in
a simulation incorporating the effects of gravitational
radiation.

There are still many issues to be resolved, but we
believe we have provided a firm foundation for future
analytic work on the evolution of a network of cosmic
strings.

\acknowledgments

We have benefited from conversations with many colleagues,
in particular Andreas Albrecht, Bruce Allen, David
Bennett, Franz Embacher, Mark Hindmarsh, Paul Shellard,
Albert Stebbins,  Neil Turok, Tanmay Vachaspati and Alex
Vilenkin.  We are grateful for the hospitality of the
Instititute  for Theoretical Physics, University of
California at Santa Barbara, where part of this work was
performed. This research was supported in part by the
National  Science Foundation under Grant No. PHY89--04035.
D.A. and E.C. are indebted to the Science and Engineering
Research Council for support.

\begin{figure}
\caption{Two contiguous segments\label{fig1}}
\end{figure}
\begin{figure}
\caption{Three contiguous segments\label{fig2}}
\end{figure}
\begin{figure}
\caption{Effect of intercommuting\label{fig3}}
\end{figure}
\begin{figure}
\caption{Position of excised loop segment\label{fig4}}
\end{figure}
\begin{figure}
\caption{Excision of a small loop\label{fig5}}
\end{figure}
\begin{figure}
\caption{Coordinates on the string world sheet\label{fig6}}
\end{figure}

\end{document}